\documentclass[12pt,a4paper]{article}
\usepackage[latin1]{inputenc}
\usepackage{amsmath}
\usepackage{mathtools}
\usepackage{setspace}
\usepackage{slashed}
\usepackage{amsfonts}
\usepackage{amssymb}
\usepackage{graphicx}
\usepackage[margin=1in]{geometry}
\usepackage{caption}
\usepackage{subcaption}
\usepackage{tikz}
\usetikzlibrary{shapes.geometric, arrows}
\usepackage[nottoc]{tocbibind}

\usepackage[titletoc]{appendix}
\usepackage{cite}
\usepackage{authblk}
\usepackage{blindtext}
\usepackage{abstract}
\title{\textbf{On the Impact of Soft Photons on $B\rightarrow K \ell^+ \ell^-$ }}
\author[1,2]{Dayanand Mishra\thanks{ dayanand@prl.res.in}}
\author[1]{Namit Mahajan\thanks{ nmahajan@prl.res.in}}
\affil[1]{Physical Research Laboratory, Ahmedabad, India.}
\affil[2]{Indian Institute of Technology, Gandhinagar, India.}
\date{}
\begin{document}
	\maketitle
	\doublespacing
	\begin{abstract}
		We give detailed results of $\mathcal{O}(\alpha)$ QED corrections (both real emission and virtual corrections) to $B\to K\ell^+\ell^-$ modes. Requiring the real emission to be gauge invariant, the structure of the contact term(s) is fixed. The calculation is done with a fictitious photon mass as the IR regulator and results are shown to be independent of it, establishing cancellation of the soft divergences. Results are presented for a choice of photon energy, $k_{max}$, and photon angle $\theta_{cut}$. The QED effects are negative, thereby reducing the rate compared to that without QED effects. Electron channels are shown to receive large corrections ($\mathcal{O}(20\%)$). Impact on lepton flavour universality ratio, $R^{\mu e}_K$ are also discussed.
	\end{abstract}
	
	\section{Introduction}
	Flavour Changing Neutral Currents (FCNCs), being both loop and CKM suppressed within the Standard Model (SM), are ideal hunting ground for physics beyond the SM. Quark level transition $b\to s \ell^+\ell^-$ have played an important role in our quest for searching for new physics as well as better understanding of the very interesting coupled dynamics of the electroweak and strong interactions governing the purely leptonic channel $B_s\to \ell^+\ell^-$ \cite{ Buras:2012ru,Hermann:2013kca,Bobeth:2013uxa,Beneke:2017vpq} and exclusive semi-leptonic channels $B\to K^{(*)}\ell^+\ell^-$ (see for example \cite{Kruger:1999xa,Ali_2000,Beneke:2004dp,Ali:2006ew,Altmannshofer:2008dz,Khodjamirian:2010vf,Beylich:2011aq,Altmannshofer:2011gn,Hambrock:2012dg,DescotesGenon:2012zf,Khodjamirian:2012rm,Jager:2012uw,Bobeth:2012vn,Descotes-Genon:2013vna,Lyon:2014hpa,Straub:2015ica}).
	 
	More and better data have hinted at some deviations \cite{Aaij:2014ora, Aaij:2017vbb, Aaij:2019wad} from the SM expectations \cite{Bordone:2016gaq} in $B\to K\ell^+\ell^-$ decays. Though not completely conclusive at this point, these deviations could be hinting at new physics just at the corner. However, such an unambiguous conclusion is somewhat masked due to hadronic uncertainties emanating from form factors as well as possible contamination from other long distance effects like tails of the charmonium resonances. \\
	The quest for precision tests of the SM via the FCNCs and searches for possible new physics have led to consideration of theoretically clean observables (sometimes also called optimised observables in specific contexts). The basic idea is to consider or construct observables - usually ratios of quantities - which are (almost) free of the hadronic uncertainties, at least in a chosen kinematic range. The decay modes $B\to K\ell^+\ell^-$ allow to test the lepton flavour universality (LFU) ie whether the decays into $\ell = e$ or $\mu$ proceed with equal strengths. Within the SM, the flavour universal coupling of the Z-boson with the leptons ensures this, of course up to the difference in the lepton masses. If the kinematical range is chosen such that the dilepton invariant mass is way larger than for either of the leptons chosen, then it is expected that the ratio of the two branching fractions is unity to a high accuracy. To this end, the following quantity is often considered as a clean test of the LFU and thus SM itself \cite{Hiller:2003js}:
	\begin{equation}
	R^{\mu e}_{K}\equiv\frac{\int_{1GeV^{2}}^{6GeV^{2}}dq^{2}\frac{d\Gamma(B^{0}\to K^{0}\mu^{+}\mu^{-})}{dq^{2}}}{\int_{1GeV^{2}}^{6GeV^{2}}dq^{2}\frac{d\Gamma(B^{0}\to K^{0}e^{+}e^{-})}{dq^{2}}}
	\end{equation}
	Within SM, this ratio is unity while experimentally it has shown deviations from this expectation\cite{Aaij:2019wad}:
	\begin{equation}
	R^{\mu e}_{K}\vert_{exp} =0.846^{+0.060\hspace{0.1cm}+0.016}_{-0.054\hspace{0.1cm}-0.014}   
	\end{equation}	
	More data and smaller errors will eventually provide possible evidence of new physics with greater confidence. In the meantime, it is important to critically analyse all possible sources of uncertainties or any other effects that may affect theoretical predictions. Theoretically, following the standard approach of operator product expansion and integrating out heavy degrees of freedom, an effective Hamiltonian is built out of relevant degrees of freedom which is then evolved down to the scale of b quark with the help of renormalization group equations (RGEs). With this set of quark level operators, the physical hadronic matrix elements are computed and it is this step that involves the introduction of the form factors. As mentioned above, some of the largest uncertainties stem from form factors and this deficiency is partly taken care of by considering observables that have very little sensitivity to form factors. The strong interaction effects, both perturbative and non-perturbative, are included via the RGEs and form factors respectively. Are there any other effects that may be relevant but have not been included explicitly? In particular, what about QED corrections? With charged particles involved, the soft photon corrections could be non-negligible and therefore should be systematically included. Such soft photon corrections have been computed for B-decays \cite{Atwood:1989em,deBoer:2018ipi,Becirevic:2009aq,Isidori:2007zt,Burdman:1994ip} and have been shown to have some impact.\\
	In the present work, we focus our attention on the modes $B\to K\ell^+\ell^-$ and compute the QED corrections. There are both virtual corrections to ${\mathcal{O}}(\alpha)$ as well as one photon emission real contributions. As is well known \cite{Yennie:1961ad,bloch,Jauch:1976ava,weinberg_1995,Hollik:1988ii,Grammer:1973db,chung,Marciano:1975de}, the sum of the two yield a finite result. These effects are important in the light of the relevance of these modes. As this work was midway, \cite{Isidori:2020acz} appeared which addresses the same issues. The results found here broadly agree with those in \cite{Isidori:2020acz}, though there are some differences. We comment on these later.
	
	\section{$B\to K\ell^+\ell^-$ without QED corrections}
	The effective Hamiltonian relevant for describing the $b\to s \ell^+\ell^-$ transition reads   \cite{Buras:1998raa,Misiak:2010zz}
	\begin{equation}
	H_{eff}=4\frac{G_{F}}{\sqrt{2}}V_{ts}^{*}V_{tb}\sum_{i=1}^{10} C_{i}(\mu)\mathcal{O}_{i}(\mu)
	\end{equation}            
	Where, $G_{F}$ is Fermi constant, $V_{ts}^{*}$ and $V_{tb}$ are CKM elements, $C_{i}$ are the Wilson coefficients and ${O}_{i}$ are the operators. $\mu$ is the scale separating long distance physics from short distance physics. $C_{i}(\mu)$ contain the information of short distance physics and can be determined using perturbation theory. ${O}_{i}(\mu)$ contain the information of long distance physics. The matrix elements of these operators are genuine non-perturbative quantities which can be parametrized in terms of form factors using Dirac and Lorentz structure. The operators most relevant for this semi-leptonic process are $O_{7}$, $O_{9}$ and $O_{10}$:
	\begin{eqnarray}
	O_{7} &=& -\frac{e}{16\pi^{2}}\frac{2 m_{b}}{q}i(\bar{s}\sigma_{\mu\nu}q^{\nu}Rb)(\bar{l}\gamma^{\mu}l) \nonumber\\
	O_{9} &=& \frac{e^{2}}{16\pi^{2}}(\bar{s}\gamma_{\mu}Lb)(\bar{l}\gamma^{\mu}l) \nonumber\\
	O_{10}& =& \frac{e^{2}}{16\pi^{2}}(\bar{s}\gamma_{\mu}Lb)(\bar{l}\gamma^{\mu}\gamma_{5}l)
	\end{eqnarray}
	The Wilson coefficients used in the calculation are: $C_{7}^{eff}=-0.319$, $C_{9}=4.228$ and $C_{10}=-4.410$. 
	$C_{9}^{eff}$ is defined by \cite{Buras} 
	\begin{eqnarray}
	C_{9}^{eff}&=&C_{9}+Y^{pert}(q^{2})\nonumber\\
	Y^{pert}(q^{2})&=&h(q^{2},m_{c})\left(\frac{4}{3}C_{1}+C_{2}+6C_{3}+60C_{5})\right)-\frac{1}{2}h(q^{2},m_{b})\Big(7C_{3}+ \frac{4}{3}C_{4}\nonumber\\
	&+&76C_{5}+\frac{64}{3}C_{6}\Big)-\frac{1}{2}h(q^{2},0)\left(C_{3}+\frac{4}{3}C_{4}+16C_{5}+\frac{64}{3}C_{6}\right)\nonumber\\
	&+&\frac{4}{3}C_{3}+\frac{64}{9}C_{5}+\frac{64}{27}C_{6}
	\end{eqnarray}
	where
	\begin{eqnarray}
	h(q^{2},m_{q})=\frac{4}{9}\ln(\frac{m_{q}^2}{\mu^{2}})+\frac{8}{27}+\frac{4}{9}(\frac{4m_{q}^{2}}{q^{2}})-\frac{4}{9}(2+\frac{4m_{q}^{2}}{q^{2}})\sqrt{\left|\frac{4m_{q}^2}{q^{2}}-1\right|}\nonumber\\
	\begin{cases}
	-\frac{i\pi}{2}+\ln{\left( \frac{1+\sqrt{1-\frac{4 m_{q}^{2}}{q^{2}}}}{\sqrt{\frac{4m_{q}^{2}}{q^{2}}}}\right)}, & \text{if} \hspace{0.1cm}\frac{4 m_{q}^{2}}{q^{2}}\leq 1\\
	\arctan\left(\frac{1}{\sqrt{\frac{4m_{q}^{2}}{q^{2}}-1}} \right), & \text{if}\hspace{0.1cm} \frac{4m_{q}^{2}}{q^{2}}< 1
	\end{cases}
	\end{eqnarray}
	The amplitude for the process $B(p_0)\to K(p_1)\ell^+(p_2)\ell^-(p_3)$ can be written in the following form \cite{Ali_2000}
	\begin{eqnarray}
	\mathcal{M}&=&\left(\frac{G_{F} \alpha |V^{*}_{ts}V_{tb}|}{2\sqrt{2}\pi}\right)\left[T_{\mu}^{1}(\bar{l}\gamma^{\mu}l)+T_{\mu}^{2}(\bar{l}\gamma^{\mu}\gamma^{5}l)
	\right]\\
	&=& \left(\frac{G_{F} \alpha |V^{*}_{ts}V_{tb}|}{2\sqrt{2}\pi}\right)(\bar{l}\Gamma^{\mu}_Al)\otimes H_{A \mu}(p,p') \nonumber
	\end{eqnarray}
	where
	\begin{equation}
	\Gamma^{\mu}_{A=1}=\gamma^{\mu},\hspace{3cm} T_{1 \mu}(p_{0},p_{1})=A'p_{\mu}+B'q_{\mu}
	\end{equation}
	and
	\begin{equation}
	\Gamma^{\mu}_{A=2}=\gamma^{\mu}\gamma^{5},\hspace{2.5cm} T_{2 \mu}(p_{0},p_{1})=C'p_{\mu}+D'q_{\mu}
	\end{equation}
	Further, we have defined the following combinations of momenta: $p_{\mu}=(p_{0}+p_{1})_{\mu}$ and $q_{\mu}=(p_{0}-p_{1})_{\mu}=(p_{2}+p_{3})_{\mu}$. In the following, the two kinematical invariants used frequently are given by $s=p^2=(p_0+p_1)^2$, and the dilepton invariant mass squared, $q^2 \equiv (p_{0}-p_{1})^{2}=(p_2+p_3)^2$.	The other factors entering the amplitude above, depending on the combinations of the Wilson coefficients ($C_7^{eff}$, $C_9^{eff}$ and $C_{10}$) and form factors ($f_+$, $f_-$ and $f_T$), are given as:
	
	\begin{eqnarray}
	A'&=& C_{9}^{eff} f_{+}(q^{2}) + \frac{2m_{b}}{m_{K}+m_{B}}C_{7}^{eff} f_{T}(q^{2}), \nonumber \\ 
	B'&= & C_{9}^{eff} f_{-}(q^{2}) - \frac{2m_{b}(m_{B}-m_{K})}{q^{2}}C_{7}^{eff} f_{T}(q^{2}), \nonumber\\
	C' &=& C_{10}f_{+}(q^{2}), \hskip 1.5cm 
	D' = C_{10}f_{-}(q^{2})
	\end{eqnarray}
	
	The non-radiative differential decay width is given by:
	\begin{equation}
	\frac{d^{2}\Gamma_{0}(B\to Kl^{+}l^{-})}{dq^{2} ds}=\frac{1}{(2\pi)^{3}}\frac{1}{32m_{B}^{3}}|M_0(B\to Kl^{+}l^{-})|^{2}
	\end{equation}
	where the explicit form of the matrix element is:
	\begin{eqnarray}
	M_0(B\to Kl^{+}l^{-}) &=& \frac{G_{F}\alpha}{2\sqrt{2}\pi}V_{ts}^{*}V_{tb} \Bigg[\Bigg( \left\lbrace C_{9}^{eff}f_{+}+C_{7}^{eff}\frac{2f_{T}m_{b}}{m_{B}+m_{k}}\right\rbrace p^{\mu} \nonumber \\
	&+& \left\lbrace C_{9}^{eff}f_{-} 
	-   C_{7}^{eff}\frac{2f_{T}m_{b}}{q^{2}}(m_{B}-m_{k})\right\rbrace q^{\mu}  \Bigg)\left(\bar{l}\gamma_{\mu}l\right) \nonumber \\
	&-& \left( C_{10}f_{+}p^{\mu}+C_{10}f_{-}q^{\mu}\right) \left(\bar{l}\gamma_{\mu}\gamma_{5}l \right) \Bigg]  
	\end{eqnarray}	
	The form factors $f_{+}(q^{2})$, $f_{-}(q^{2})$ and $f_{T}(q^{2})$ are parametrized as \cite{Bobeth:2011nj,Khodjamirian:2010vf}
	\begin{align*}
	f_{i}(q^{2})=\frac{f_{i}(0)}{1-\frac{q^{2}}{m_{res,i}^{2}}}\left[ 1+b_{1i}\left( z(q^{2})-z(0)+\frac{1}{2}\left(z^{2}(q^{2})-z^{2}(0)\right)\right)\right]
	\end{align*}
	where, $i= +,\hspace{0.1cm}0,\hspace{0.1cm}T$;
	\begin{align*}
	f_{-}=\left(f_{0}-f_{+} \right)\frac{m_{B}^{2}-m_{K}^{2}}{q^{2}}
	\end{align*}
	$z(q^{2})$ is defined by
	\begin{eqnarray}
	z(q^{2})&=\frac{\sqrt{\tau_{+}-q^2}-\sqrt{\tau_{+}-\tau_{0}}}{\sqrt{\tau_{+}-q^2}+\sqrt{\tau_{+}-\tau_{0}}}
	\end{eqnarray}
	where
	\begin{eqnarray}
	\tau_{0}&=\sqrt{\tau_{+}}\left(\sqrt{\tau_{+}}-\sqrt{\tau_{+}-\tau_{-}} \right), \hskip 0.5cm
	\tau_{\pm}&=\left(m_{B}\pm m_{K}\right)^{2} 
	\end{eqnarray}
	with $f_{i}(0) = \{0.34,\hspace{0.1cm} 0.34,\hspace{0.1cm} 0.39\}$, $b_{1i} = \{-2.1,\hspace{0.1cm}-4.3,\hspace{0.1cm}-2.2\}$ and $m_{res\hspace{0.1cm}i} = \{5.83,\hspace{0.1cm}5.37,\hspace{0.1cm}5.41\}$ for $i=( +,\hspace{0.1cm}0,\hspace{0.1cm}T)$, respectively. 
	
	\section{QED Corrections}	
	We now consider QED corrections. Fig. (1) shows the photon emission diagrams ($\textcolor{red}{\times}$ denotes possible places from where a photon can be emitted, including $B$ and $K$ legs when charged and computed assuming the mesons to be point like and employing scalar QED). The diagram (b) is the so called {\it contact term} (CT) and arises due to the gauge invariance of QED. In \cite{Isidori:2020acz}, a mesonic level Lagrangian is assumed and following the minimal coupling prescription, these are computed. Our approach is different from theirs and is discussed in detail below. 
	\begin{figure}[h]
		\begin{subfigure}{.4\textwidth}
			\centering
			\includegraphics[width=0.7\linewidth]{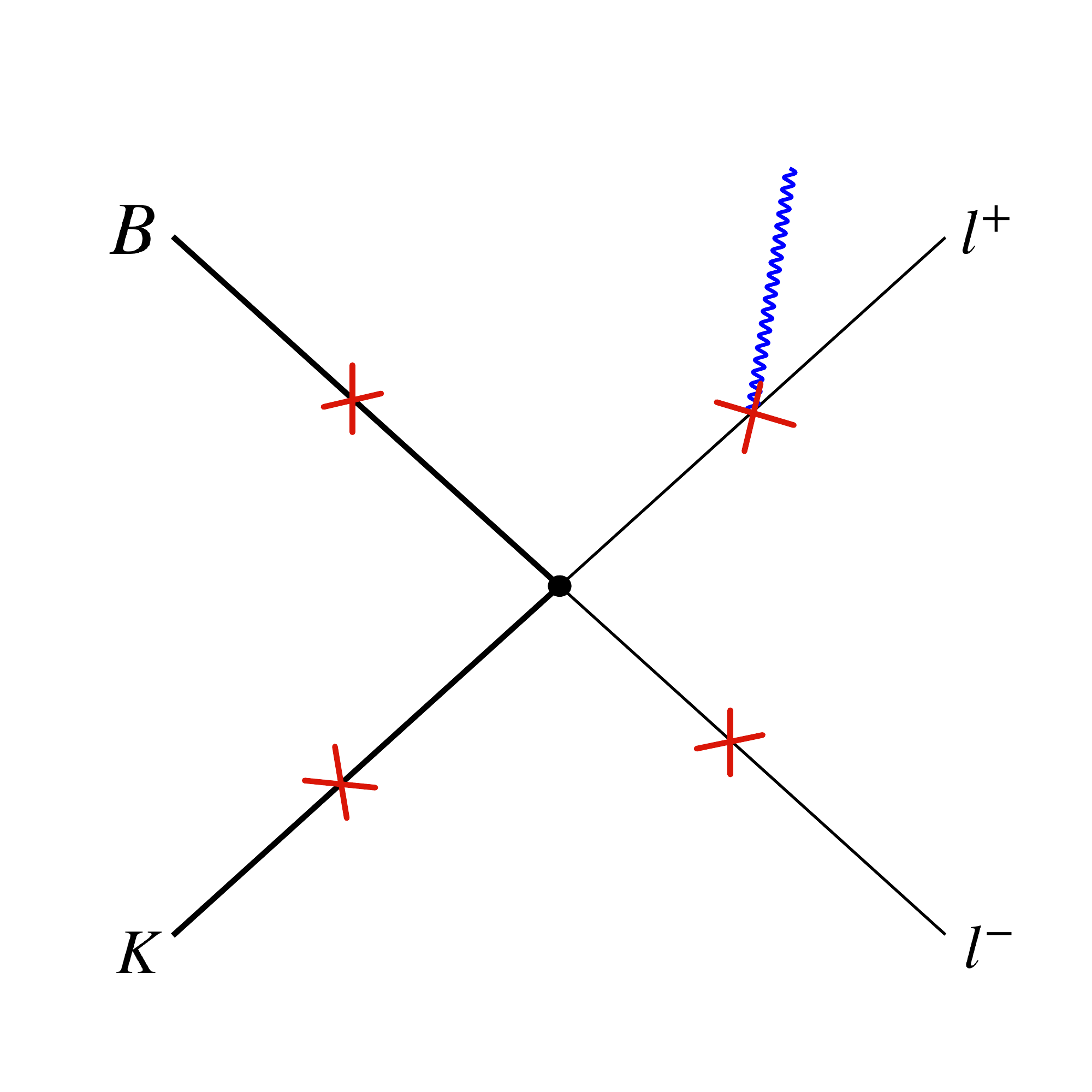}
			\caption{\textcolor{red}{X} :photon emission}
		\end{subfigure}%
		\begin{subfigure}{.4\textwidth}
			\centering
			\includegraphics[width=0.7\linewidth]{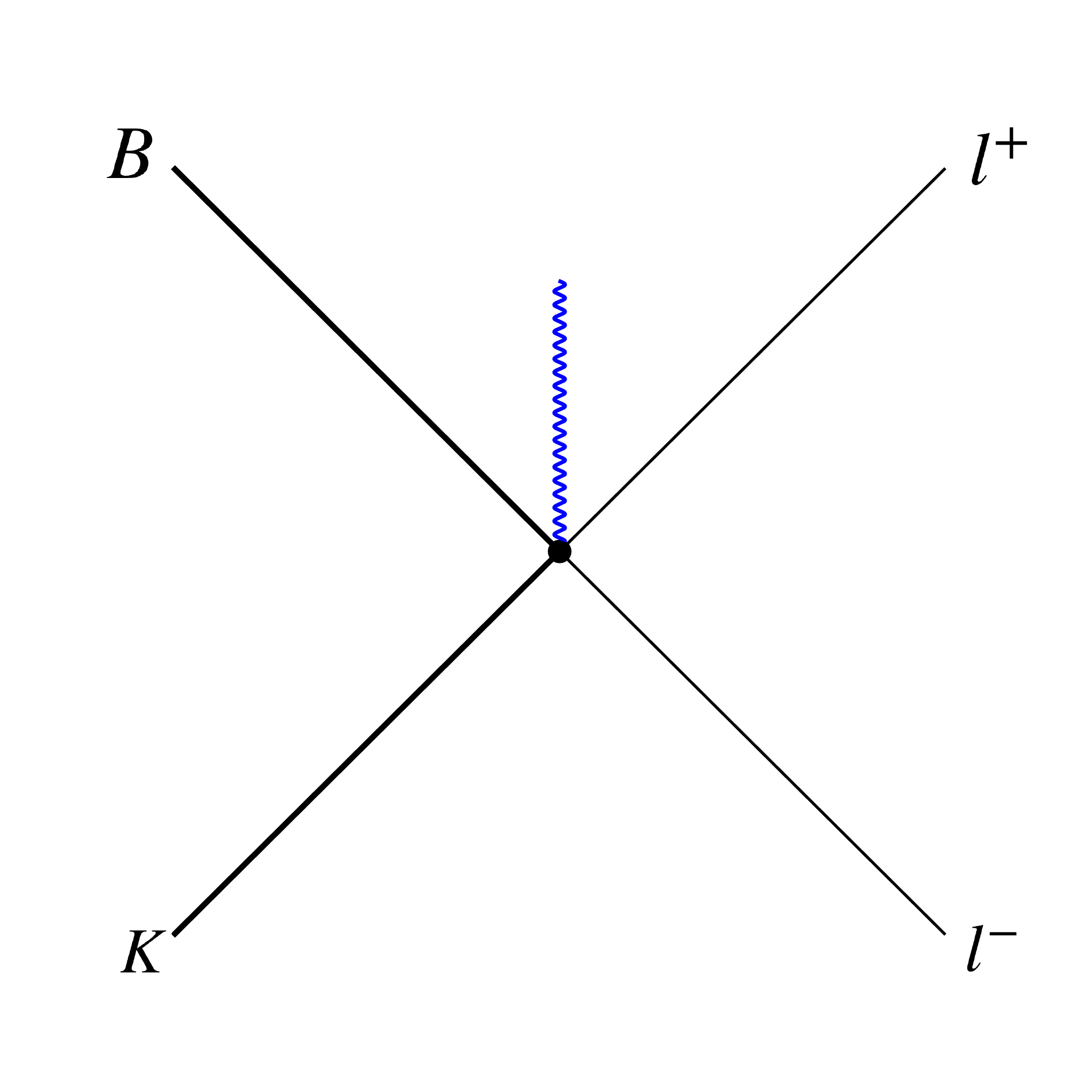}
			\caption{Contact term}
		\end{subfigure}%
		
		\caption{Representative diagram for real photon emission}
	\end{figure} 
	Fig. (2) shows some of the representative diagrams that contribute to the virtual corrections. As is clearly shown, the diagrams which involve photon from the contact term are also included to ensure cancellation of the infrared divergences and having a gauge invariant result.
	\begin{figure}[h]
		\begin{subfigure}{.25\textwidth}
			\centering
			\includegraphics[width=1.\linewidth]{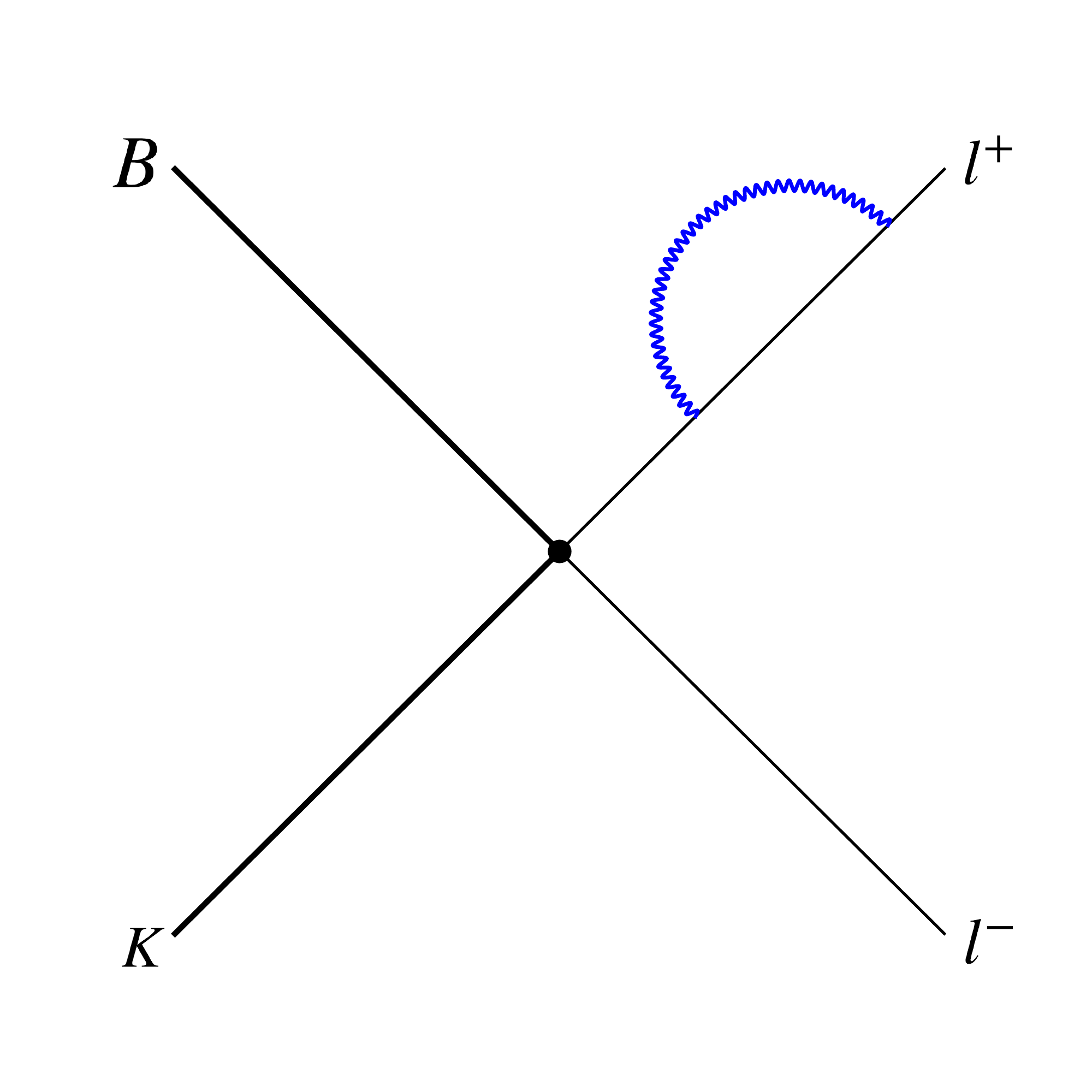}
			\caption{}
		\end{subfigure}%
		\begin{subfigure}{.25\textwidth}
			\centering
			\includegraphics[width=1.\linewidth]{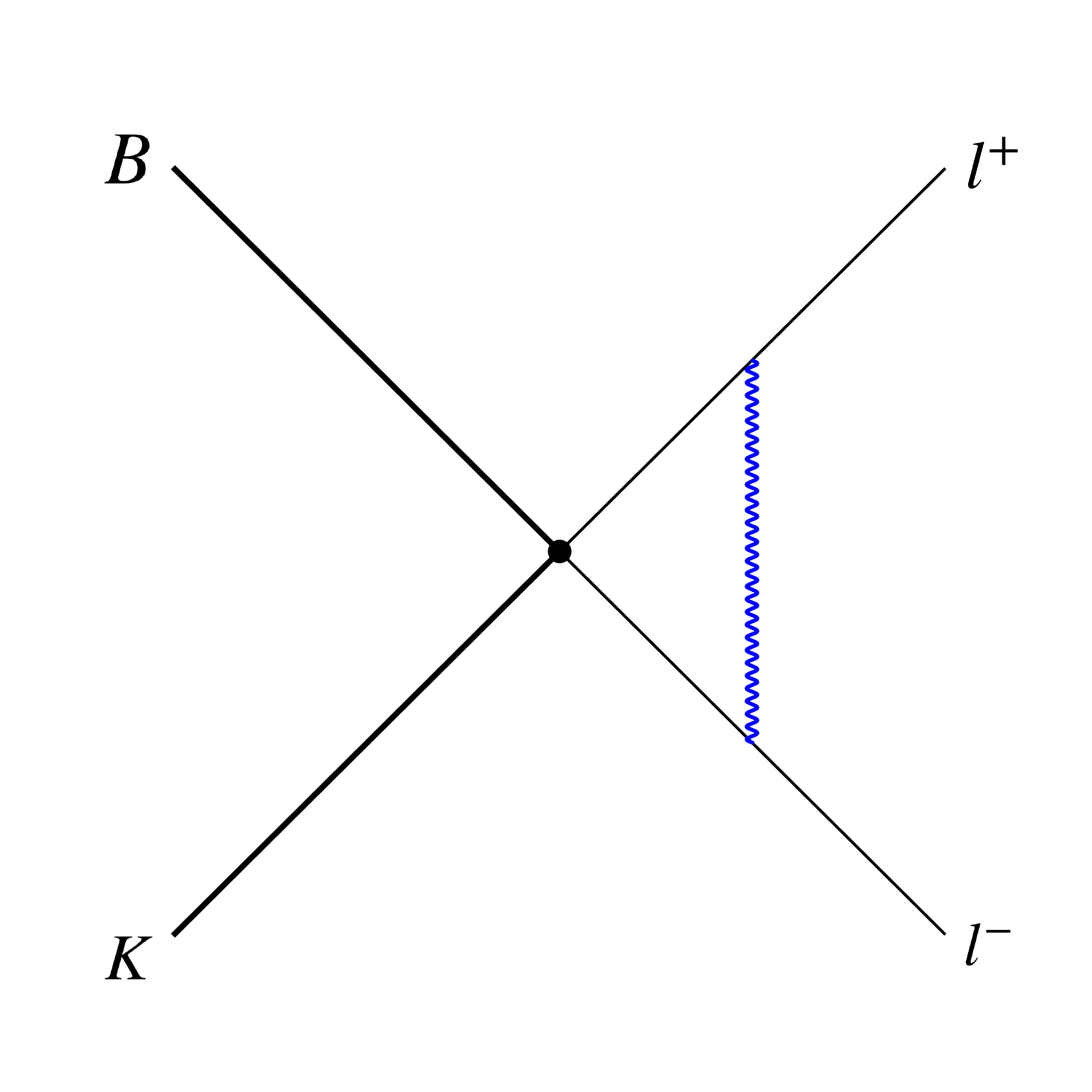}
			\caption{}
		\end{subfigure}%
		\begin{subfigure}{.25\textwidth}
			\centering
			\includegraphics[width=1.\linewidth]{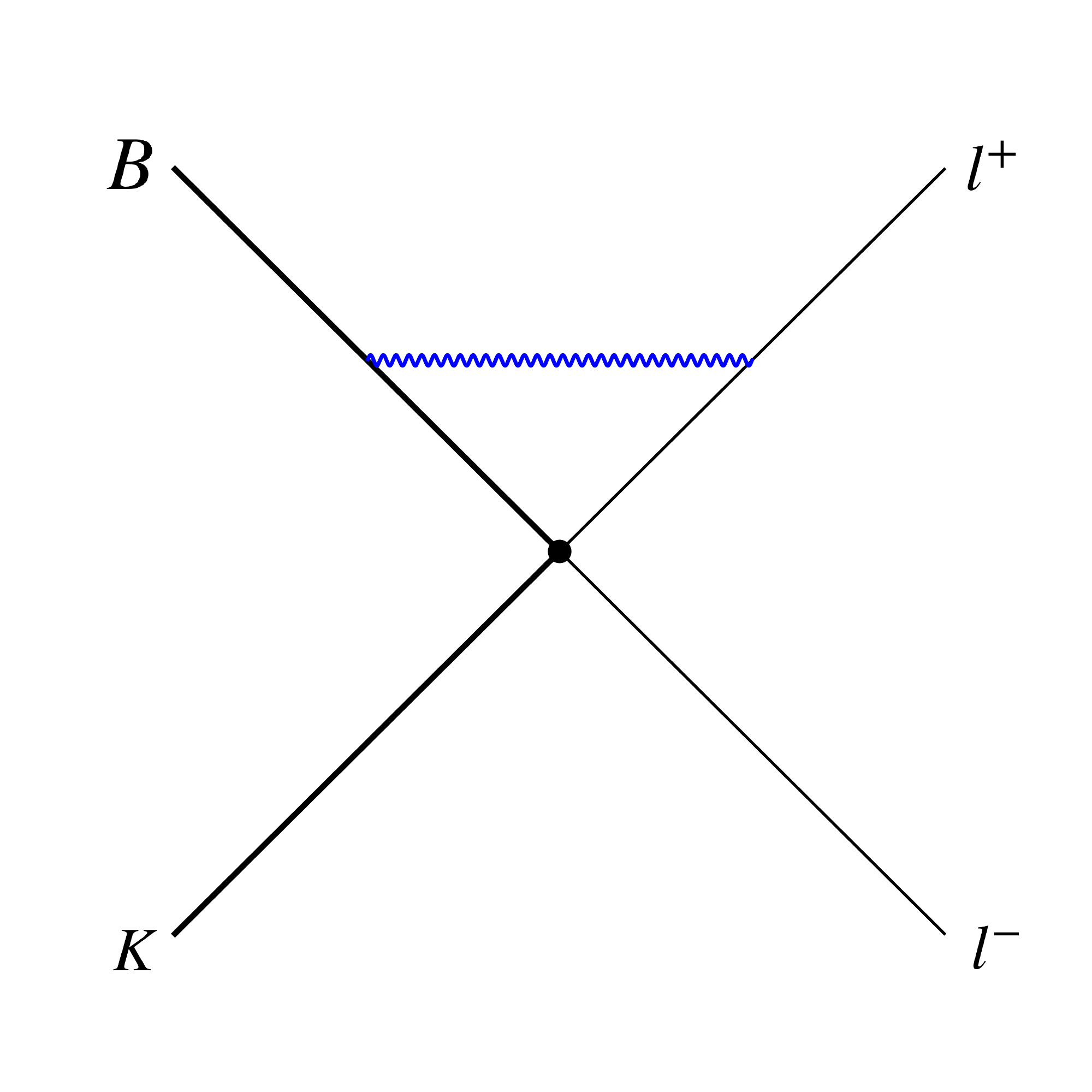}
			\caption{}
		\end{subfigure}%
		\begin{subfigure}{.25\textwidth}
			\centering
			\includegraphics[width=1.\linewidth]{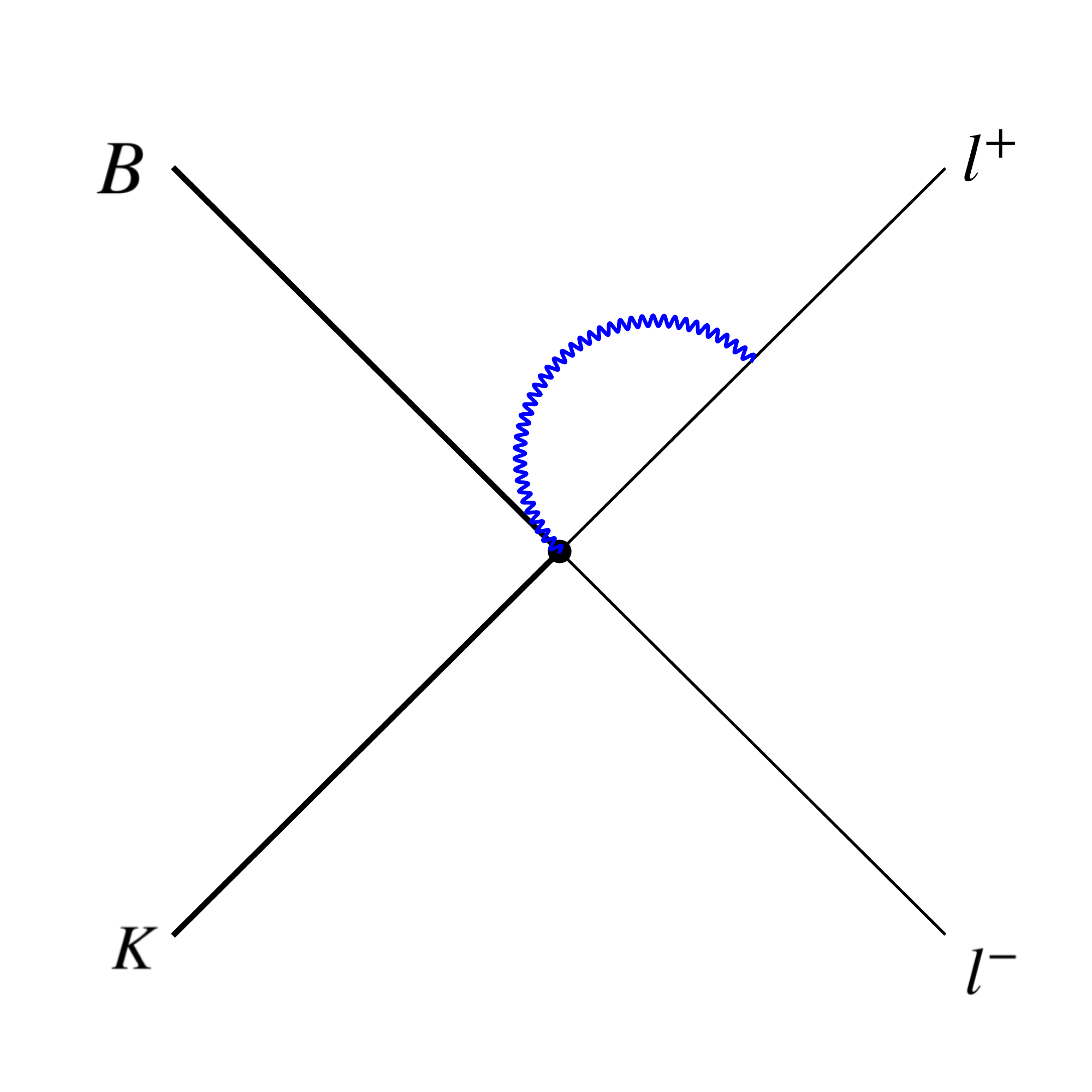}
			\caption{}
		\end{subfigure}%
		\caption{Representative diagrams contributing to virtual corrections}
	\end{figure}  
	The photon momentum is denoted by $k$ in the calculations below and the polarization vector is denoted by $\epsilon_{\alpha}(k)$. Consider the general case where the charges of the $B$ and $K$ meson are denoted by $Q_B$ and $Q_K$. Since we are only interested in lepton number conserving processes, eventually we impose $Q_B = Q_K$, but for the time being we keep them general.
	\subsection{Contact Term}
	Before explicitly computing the virtual corrections and the real emission contributions, it is important to fix the contact term. To this end, consider photon emission from different legs. The process under consideration thus is: $B(p_0)\to K(p_1)\ell^+(p_2)\ell^-(p_3)\gamma(k)$. For the mesons, employing scalar QED, the matrix element for photon emission from the external legs, written in terms of quantities for the non-radiative decay, reads 
	\begin{eqnarray}
	\tilde{M} &=& -e \epsilon_{\alpha}(k)\bar{u}(p_{2})\Gamma^{\mu}_{A}\frac{(\slashed{p}_{3}+\slashed{k})-m_{l}}{2p_{3}.k}\gamma^{\alpha}v(p_{3}) \otimes H_{A \mu}(p_{0},p_{1}) \nonumber \\
	&+& e \epsilon_{\alpha}(k)\bar{u}(p_{2})\gamma^{\alpha}\frac{(\slashed{p}_{2}+\slashed{k})+m_{l}}{2p_{2}.k}\Gamma^{\mu}_{A}v(p_{3}) \otimes H_{A \mu}(p_{0},p_{1})  \nonumber\\ 
	&+& e Q_{B} \epsilon_{\alpha}(k)\frac{2p_{0}^{\alpha}}{2p_{0}.k}\bar{u}(p_{2})\Gamma^{\mu}_{A}v(p_{3}) \otimes H_{A \mu}(p_{0}-k,p_{1})\nonumber\\
	&-& e Q_{K} \epsilon_{\alpha}(k)\frac{2p_{1}^{\alpha}}{2p_{1}.k}\bar{u}(p_{2})\Gamma^{\mu}_{A}v(p_{3}) \otimes H_{A \mu}(p_{0},p_{1}+k)
	\end{eqnarray}
	It is clear from the above equation that when the photon is emitted from one of the leptons, the momentum dependence of the hadronic part $H_{A \mu}$ remains the same as in non-radiative decay while the dependence is appropriately modified in case of emission from the meson legs. Given the explicit structure of $H_{A \mu}$, the above can be written in a more explicit and convenient form: 
	\begin{equation}
	\tilde{M}=M_{0}e\epsilon_{\alpha} \sum_{i}\frac{Q_{i}\eta_{i}p_{i}^{\alpha}}{p_{i}.k} + M'(k)
	\end{equation}
	where, $p_i$ and $Q_{i}$ are the momenta and charges of different particles respectively, and $\eta_{i}$ are $+(-)$ for outgoing (incoming) particle \cite{weinberg_1995}. $M_0$ is the amplitude for the process without the photon emission. The first term above is nothing but the Low's soft photon amplitude: $M(a\to b \gamma(k))\vert_{k\to 0} = S\otimes M(a\to b)$, where $S$ is the universal soft function given by the quantity multiplying $M_0$. It can be explicitly verified that the Low's term is gauge invariant by itself. This piece is easy to compute as the hadronic contribution is same as that in the case of non-radiative amplitude.
	The remainder is written as $M'(k)$ and is the non-infrared contribution ie unlike the Low's term which is $\mathcal{O}(1/k)$, the terms in $M'(k)$ are $\mathcal{O}(k)$ and higher. We now turn to $M'(k)$. 
	
	$M'(k)$ consists two contributions: $M'_{lept}$, arising from the emission from the leptons and $M'_{mes}$ from the mesons. These are given by:
	\begin{equation}
	M'_{lept} = e\epsilon_{\alpha}(k)\left[\bar{u}(p_{2})\gamma^{\alpha}\frac{\slashed{k}}{2p_{2}.k}\Gamma^{\mu}_{A}v(p_{3}) -\bar{u}(p_{2})\Gamma^{\mu}_{A}\frac{\slashed{k}}{2p_{3}.k}\gamma^{\alpha}v(p_{3})\right]\otimes H_{A \mu}(p_{0},p_{1})
	\end{equation}
	and 
	\begin{equation}
	M'_{mes} = -e \epsilon_{\alpha}(k)\,\left[ Q_{B}\alpha_{A}\frac{2p_{0}^{\alpha}}{2p_{0}.k}+Q_{K}\beta_{A}\frac{2p_{1}^{\alpha}}{2p_{1}.k}\right]\, \bar{u}(p_{2})\Gamma^{\mu}_{A}v(p_{3})\, k_{\mu}
	\end{equation}
	$\alpha_{A}=A'+B'$ or $C'+D'$ and $\beta_{A}=A'-B'$ or $C'-D'$ for $A=1,2$. In obtaining this form, we have made use of the fact that the general structure of $H_{A \mu}$ has the form
	\begin{equation}
	H_{A \mu}(p,q) = X_A \, p_{\mu} + Y_A \, q_{\mu}
	\end{equation}
	Thus, it is straightforward to incorporate the appropriate shifts in the momentum dependence due to an additional photon being emitted, and one immediately arrives at the form above. Therefore, the total contribution beyond Low's soft photon contribution  
	\begin{equation}
	M'(k) = M'_{lept} + M'_{mes}
	\end{equation}
	As mentioned above, Low's contribution is automatically gauge invariant. Let us now check the gauge invariance of $M'(k)$ by making the replacement: $\epsilon_{\alpha} \rightarrow k_{\alpha}$. This replacement in $M'_{lept}$ yields zero. Therefore this subset is gauge invariant by itself. \\
	Turning now to $M'_{mes}$, the replacement yields
	\begin{equation}
	M'_{mes}|_{\epsilon_{\alpha}\to k_{\alpha}}=-e(Q_{B}+Q_{K})\xi_{A} k_{\mu}\left[\bar{u}(p_{2})\Gamma^{\mu}_{A} v(p_{3}) \right]
	\end{equation}
	Here, $\xi_{A}={A'(q^{2})\,(C'(q^{2})})$ for $A = 1 (2)$.\\
	This is the extent by which gauge invariance is violated. Therefore, negative of this quantity is the contribution that should be there to ensure gauge invariance of the full amplitude. Moreover, this has the form of a contact term, and should be added to the amplitude $M'$ to ensure a gauge invariant result. Rewriting this as an additional term in the effective Hamiltonian at the hadronic level (the result for $\epsilon\to k$, therefore flipping back to $\epsilon$)
	\begin{eqnarray}
	\mathcal{H}_{eff}^{CT}&=i e\xi_{A}(Q_{B}+Q_{K}) \left[\bar{u}(p_{2})\Gamma^{\alpha}_{A} v(p_{3}) \right]A_{\alpha}
	\end{eqnarray}
	This then provides the contact term (Fig.(1b)) and will contribute both to the real emission and virtual corrections and is $\mathcal{O}(k)$. The contact term is proportional to the sum of the $Q_B$ and $Q_K$, and clearly it plays no role when $B$ and $K$ are neutral. Our way of determining the contact terms is very different from that adopted in \cite{Isidori:2020acz}. There, a mesonic level Lagrangian with specific operator structure is assumed and then following the minimal coupling prescription, $\partial_{\mu}\rightarrow \partial_{\mu} - ie A_{\mu}$, the required contact terms are obtained. The contact terms obtained here include effects due to all operators that contribute to this process. This leads to some difference in the numerical values of the corrections compared to that in \cite{Isidori:2020acz}, but as we see later, there is in general good agreement between the two results. 
	
	After having fixed the contact term, ensuring the requirement of gauge invariance of the the full amplitude, we now turn to computing the $\mathcal{O}(\alpha)$ corrections: real photon emission rate and virtual corrections to the non-radiative amplitude to that order and then squaring the amplitude retaining up to the interference terms between the lowest order and $\mathcal{O}(\alpha)$ terms. Finally, to obtain the rate to be compared with experimentally observed one, these two are incoherently added. We closely follow \cite{Yennie:1961ad} in our explicit calculations. In particular, photon is imparted a small mass $m_{\gamma} = \lambda$. This takes care of the IR divergences. The loop integrals are regularized using Dimensional Regularization.
	
	\subsection{Real Photon Emission}
	The total contribution to the real photon emission amplitude
	$B(p_0)\to K(p_1)\ell^+(p_2)\ell^-(p_3)\gamma(k)$ will then be sum of Low's IR terms, $M'(k)$ with the contribution from the contact term included properly. At the decay rate level, one then writes:
	\begin{equation}
	d\Gamma_{real} = d\Gamma_{0}(1+2 \alpha \tilde{B})  + d\Gamma'
	\end{equation}
	where, $d\Gamma_{0}$ is the non-radiative decay rate and the quantity $\tilde{B}$ given below captures the IR contribution stemming from the Low's term. 
	\begin{equation}
	\tilde{B}=\frac{1}{8\pi^2}\int \frac{d^{3}k}{\sqrt{k^2+\lambda^{2}}}\left( \sum_{i}\frac{Q_{i}\eta_{i}p_{i}^{\alpha}}{p_{i}.k}\right)^{2}
	\end{equation}
	where one can easily see the universal soft factor $S = \epsilon_{\alpha}(k) \tilde{T}^{\alpha}(k)$ for the present case
	\begin{align*}
	\tilde{T}^{\alpha}(k)=-\frac{2p^{\alpha}_{i }\eta_{i}}{2k.p_{i}\eta_{i}}+\frac{2p^{\alpha}_{j}\eta_{j}}{2k.p_{j}\eta_{j}} 
	\end{align*}
	Using charge conservation, $\sum_{i}Q_{i}\eta_{i}=0$,
	$\tilde{B}$ can be rewritten as
	\begin{equation}
	\tilde{B}=\frac{1}{8\pi^2}\int_{0}^{k_{max}} \frac{d^{3}k}{\sqrt{k^2+\lambda^{2}}} \sum_{i\neq j, i<j}Q_{i}Q_{j}\eta_{i}\eta_{j}\left(\frac{p_{i}^{\alpha}}{p_{i}.k}-\frac{p_{j}^{\alpha}}{p_{j}.k} \right)^{2}
	\end{equation}
	where $k_{max}$ has been introduced explicitly. Only photons below $k_{max}$ can't be observed experimentally and the theoretical rate will thus depend on the value of $k_{max}$. As seen from the equation above, the indices $i$ and $j$ then take values $i=1,2,3$ and $j=2,3,4$, where the numbers $1,2,3,\hspace{0.1cm}\text{and}\hspace{0.1cm}4$ represent particles $B,K,\ell^+, \hspace{0.1cm} \text{and} \hspace{0.1cm}\ell^-$ respectively.
	
	For the charged $B$ meson decay, there will be six terms in $\tilde{B}$, 
	\begin{align*}
	\tilde{B}= \tilde{B}_{BK}+ \tilde{B}_{Bl^+}+\tilde{B}_{Bl^-}+\tilde{B}_{Kl^+}+\tilde{B}_{Kl^-}+\tilde{B}_{l^+l^-}
	\end{align*} 
	After integrating over $k$, $\tilde{B}$ reads (Appendix A lists the integrals encountered during the calculation),
	\begin{equation}
	\tilde{B_{ij}}=\frac{Q_{i}Q_{j}\eta_{i}\eta_{j}}{2\pi}\left\lbrace \ln\left( \frac{k_{max}^{2}m_{i}m_{j}}{\lambda^{2}E_{i}E_{j}}\right)-\frac{p_{i}.p_{j}}{2}\left[ \int_{-1}^{1}\frac{dx}{p_{x}^{2}}\ln\left(\frac{k_{max}^{2}}{E_{x}^{2}} \right)+\int_{-1}^{1}\frac{dx}{p_{x}^{2}}\ln\left(\frac{p_{x}^{2}}{\lambda^{2}} \right) \right]  \right\rbrace
	\end{equation}
	where we have defined following combinations which are convenient parameterizations while evaluating the integrals like above:
	\begin{eqnarray}
	2p_{x} &=& (1+x)p_{i}+(1-x)p_{j} \nonumber\\
	2E_{x}&=&(1+x)E_{i}+(1-x)E_{j} \nonumber \\
	2p'_{x} &=&(1+x)p_{i}\eta_{i}-(1-x)p_{j}\eta_{j}
	\end{eqnarray}
	Here $p_{i,j}$ are the four momenta while $E_{i,j}$ are the energies of the particles. These lead to
	\begin{eqnarray}
	p_{x}^{2} &=& (1+x)^{2}p_{i}^{2}+(1-x)^{2}p_{j}^{2}+2(1-x^{2})p_{i}.p_{j} \nonumber \\
	p_{x}^{'2} &=& (1+x)^{2}p_{i}^{2}+(1-x)^{2}p_{j}^{2}-2(1-x^{2})p_{i}.p_{j}\eta_{i}\eta_{j}
	\end{eqnarray}
	The non-IR contribution includes terms beyond the Low's term in the amplitude, at order $\mathcal{O}(k)$ and higher. The remaining terms in the decay rate are the square of these non-IR terms and the interference terms between the IR and non-IR terms. The contribution from the square of the non-IR terms is found to be negligible and is not considered in the analysis. On the other hand, the interference terms produce relevant contribution. These interference terms depend on $\theta$, the angle between negatively charged lepton and photon. We observe (as shown below) that the correction factor as defined later, denoted $\Delta^{i}$ ($i=e$) $\mu$ below, is very sensitive to lower angular cut $\theta_{cut}$ for $i=e$ due to the smallness of electron mass, while for the chosen values of $k_{max}$, there is not much effect when $i=\mu$. 
	
	\subsection{Virtual Photon Corrections}
	There are three kind of diagrams contributing to virtual corrections: (a) photon starting and ending at the same charged particle leg (Fig.(2a)); (b) photon line between two different charged particles (Fig.(2b,2c)); (c) photon from the contact term ending on a charged particle leg (Fig.(2d)).
	
	We first consider the set of diagrams arising due to the contact term. Specifically, considering the case when the photon from the effective contact term vertex attaches to the lepton leg. This contribution gets cancelled by an equally opposite one coming diagram where the photon attaches to the anti-lepton leg. The other two diagrams with the photon from the contact vertex ending at either the $B$ or $K$ leg can be evaluated in a straightforward manner. These lead to UV divergences and a finite part ($M_{CT}$). These UV divergences would require extra higher dimensional operators to be cancelled or absorbed systematically. It is worth noting that the way we arrived at the contact term was by requiring the real emission amplitude to be gauge invariant. This amplitude is $\mathcal{O}(e)$, and therefore the contact term that one gets is of this order only, ie with one photon. Moreover, it is possible that one misses out on terms that vanish for on-shell photon but can contribute to virtual corrections. Also when evaluating the virtual corrections, there is an extra factor of $e$ and this correction if $\mathcal{O}(e^2)$. From an effective theory point of view, there can be, or rather will be, operators corresponding to a term with leptons, $B$ and $K$ mesons and two photons like the one in Fig.(3). These would lead to diagrams of the type shown on the right in Fig.(3). In dimensional regularization, scale less integrals would simply be zero. One would also require to include other higher dimensional operators at $\mathcal{O}(e)$, including the derivative operators, to the given order for consistency.
	\begin{figure}[h]
		\begin{subfigure}{.4\textwidth}
			\centering
			\includegraphics[width=0.7\linewidth]{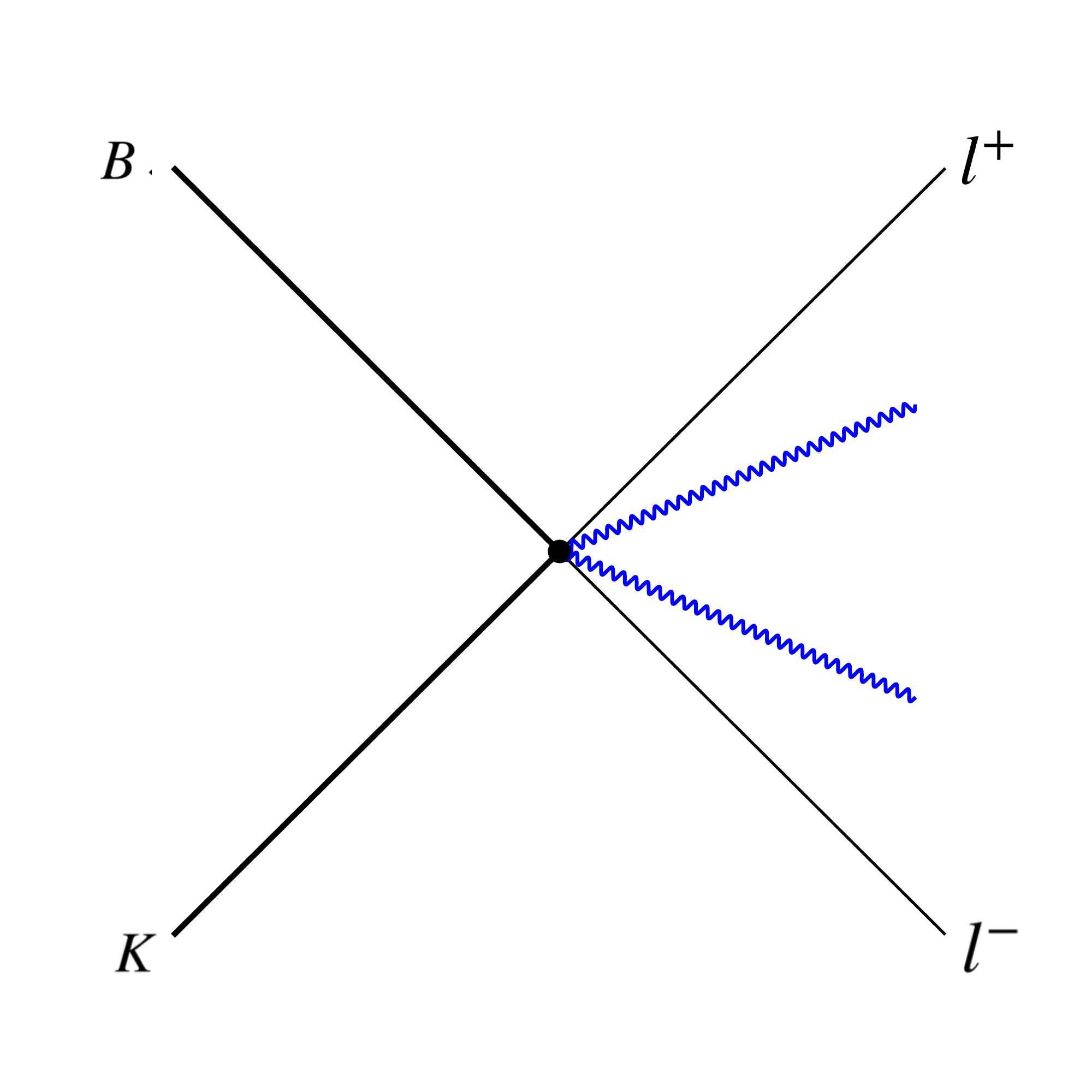}
		\end{subfigure}%
		\hskip 2.0cm
		\begin{subfigure}{.4\textwidth}
			\centering
			\includegraphics[width=0.65\linewidth]{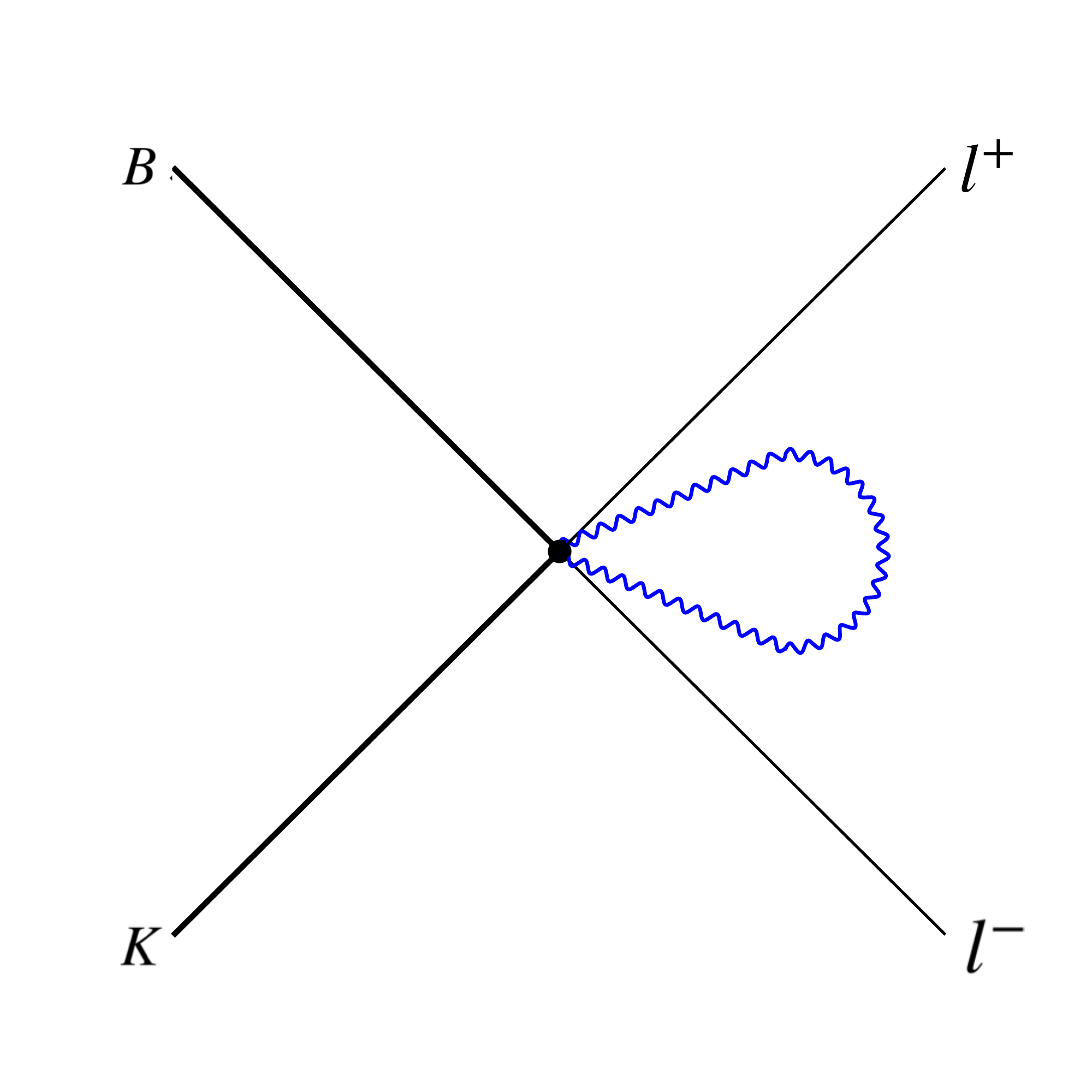}
		\end{subfigure}%
		\caption{Two photon Contact Term. (Left) real emission (Right) Virtual correction}
	\end{figure} 
	One may now start with the effective Hamiltonian, including the one photon contact term, and then require the two photon emission amplitude to be gauge invariant, thereby fixing the two photon contact term plus possibly more new terms, which would actually correspond to some higher dimensional operators. However, again there is no guarantee that the terms will get completely fixed since one again is requiring on-shellness of the two photons. Or as the suggested prescription in \cite{Isidori:2020acz}, consider higher dimensional derivative operators and using minimal coupling, generate required terms. However, recalling that minimal coupling prescription may be ambiguous \cite{Jenkins:2013fya}, utmost care should be taken to fix the structure of such terms and also to keep in mind that there could be more than one structure for such terms.\footnote{That there will be need to go beyond minimal coupling is also pointed out in \cite{Isidori:2020acz}.} Perhaps the right way would be to start with the quark level operators and compute the following matrix element say within sum rule approach
	\begin{equation*}
	\langle K(p_1)\ell^+\ell^-\gamma(k)\vert (\bar{\ell}\Gamma_{\mu}\ell)\, (\bar{s}\Gamma'_{\mu} b)\vert B(p_0)\rangle \propto \epsilon^{\alpha}\int d^4x e^{i k.x}\langle K(p_1)\ell^+\ell^-\vert T[j_{\alpha}^{em}(x) (\bar{\ell}\Gamma_{\mu}\ell)\, (\bar{s}\Gamma'_{\mu}b) (0)]\vert B(p_0)\rangle
	\end{equation*}
	This will in general have two type of terms: (i) photon emission from leptons times the $B$ to $K$ matrix element, (ii) photon being emitted from the hadronic system. Using the QED Ward identity, the general structure of the hadronic tensor can then be fixed and finally, the new form factors can be evaluated, or at least estimated in the factorization approximation. The procedure can then be repeated for two photon emission case. It is evident that it'll keep becoming rather daunting to evaluate newer matrix elements. This will be taken up separately, comparing the results of different approaches.\\
	Given these issues, which are certainly beyond the scope of the present work, we choose to simply discard the UV divergences encountered when evaluating the diagrams involving the contact term, and include the finite parts in our calculation. Explicit numerical checks show that these finite parts are rather small and do not affect the results at the level of precision the whole calculation is being done.	
	Evaluating the rest of the diagrams, one then finds
	\begin{equation}
	M_{\text{virtual}}=M_{0}\left[ 1+\alpha B+\frac{\alpha}{2\pi}\right] + M_{CT}
	\end{equation} 
	In this equation, the last term in the parenthesis is the magnetic moment like term, which is free of divergences. The quantity denoted as $B$ contains contributions from the self energy and vertex corrections and reads
	\begin{equation}
	B=\frac{i}{8\pi^{3}}\int d^{4}k\frac{1}{(k^{2}-\lambda^{2}+i\epsilon)} \left[ \sum_{i=1}^{4}\frac{Q_{i}^{2}(2p_{i}-k)^{2}}{(k^{2}-2k.p_{i})^{2}}-2\sum_{i\neq j, i<j}\frac{Q_{i}Q_{j}\eta_{i}\eta_{j}(2p_{i}\eta_{i}-k).(2p_{j}\eta_{j}+k)}{(k^{2}-2k.p_{i}\eta_{i})(k^{2}+2k.p_{j}\eta_{j})}\right]
	\end{equation}	 
	Using charge conservation $\sum_{i}Q_{i}\eta_{i}=0$, $B$ can be written as
	\begin{equation}
	B=\frac{-i}{8\pi^{3}}\int d^{4}k\frac{1}{(k^{2}-\lambda^{2}+i\epsilon)}\sum_{i\neq j,i<j}Q_{i}Q_{j}\eta_{i}\eta_{j}\left(\frac{2p^{\alpha}_{i }\eta_{i}-k^{\alpha}}{k^{2}-2k.p_{i}\eta_{i}}+\frac{2p^{\alpha}_{j}\eta_{j}+k^{\alpha}}{k^{2}+2k.p_{j}\eta_{j}} \right)^{2}
	\end{equation}
	where like for the real emission  $i=1,2,3$ and $j=2,3,4$, and the numbers $1,2,3,\hspace{0.1cm}\text{and}\hspace{0.1cm}4$ represent particles $B,K,\ell^+, \hspace{0.1cm} \text{and} \hspace{0.1cm}\ell^-$ respectively.
	Now, for $B\to K \ell^+ \ell^-$, when both the mesons and the leptons are charged, a total of six diagrams will contribute to $B$ 
	\begin{align*}
	B=B_{BK}+B_{Bl^+}+B_{Bl^-}+B_{Kl^+}+B_{Kl^-}+B_{l^+l^-}
	\end{align*}
	After integrating over $k$ and employing Dimensional Regularization, the general structure of $B_{ij}$ is calculated to be
	\begin{equation}
	B_{ij}=\frac{-1}{2\pi}Q_{i}Q_{j}\eta_{i}\eta_{j}\left[ln(\frac{m_{i}m_{j}}{\lambda^{2}})+\frac{1}{4}\int_{-1}^{1}dx ln(\frac{p_{x}^{'2}}{m_{i}m_{j}})+\frac{p_{i}.p_{j}\eta_{i}\eta_{j}}{2}\int_{-1}^{1}\frac{dx}{p_{x}^{'2}}ln(\frac{p_{x}^{'2}}{\lambda^{2}})\right]
	\end{equation}
	Relevant integrals encountered in the intermediate steps are collected in appendix B. 
	
	\subsection{Sommerfeld factor}
	We have also considered the Sommerfeld enhancement factor (Coulomb factor), arising due to the difference in the scattering in the presence of potential versus the absence of the potential \cite{sommerfeld}. This correction is a multiplicative factor given by
	
	\begin{equation}
	\Omega_{c}=\frac{-2\pi \alpha}{\beta_{ij}}\frac{1}{e^{\frac{-2\pi \alpha}{\beta_{ij}}}-1}
	\end{equation}
	where,
	\begin{equation}
	\beta_{ij}=\sqrt{1-\frac{m_{i}^{2}m_{j}^{2}}{(p_{i}.p_{j})^{2}}}
	\end{equation}
	with $\beta_{ij}$ representing relative velocity between $i^{th}$ and $j^{th}$ particle. 
	
	\subsection{Total $\mathcal{O}(\alpha)$ QED corrections}
	We are now in a position to compute the decay rate to $\mathcal{O}(\alpha)$: 
	\begin{equation}
	d\Gamma_{real} = d\Gamma_{0}\left(1+2 \alpha \tilde{B} + 2 \alpha B + \frac{\alpha}{\pi}\right) \Omega_{c}
	\end{equation}
	Both $\tilde{B}$ and $B$ (or equivalently $\tilde{B}_{ij}$ and $B_{ij}$) depend on the fictitious photon mass $\lambda$, which was introduced to regulate the IR divergences. The result for the physical rate above should be independent of $\lambda$. Defining
	$\mathcal{H}_{ij}=B_{ij}+\tilde{B}_{ij}$ given as
	\begin{eqnarray}
	\mathcal{H}_{ij} &=& \frac{-Q_{i}Q_{j}\eta_{i}\eta_{j}}{2\pi}\left[-\ln\left( \frac{k_{max}^{2}}{E_{i}E_{j}}\right)+\frac{1}{4}\int_{-1}^{1}dx \ln\left( \frac{p_{x}^{'2}}{m_{i}m_{j}}\right)\nonumber \right.\\ &+&\left. \frac{p_{i}.p_{j}\eta_{i}\eta_{j}}{2}\int_{-1}^{1}\frac{dx}{p_{x}^{'2}}\ln\left( \frac{p_{x}^{'2}}{\lambda^{2}}\right) 
	+ \frac{p_{i}.p_{j}}{2} \int_{-1}^{1}\frac{dx}{p_{x}^{2}}\ln\left(\frac{k_{max}^{2}}{E_{x}^{2}} \right)\nonumber \right.\\
	&+&\left. \frac{p_{i}.p_{j}}{2}\int_{-1}^{1}\frac{dx}{p_{x}^{2}}\ln\left(\frac{p_{x}^{2}}{\lambda^{2}} \right)\right]
	\end{eqnarray}
	Recalling\\
	$p_{x}^{'2}=(1+x)^{2}p_{i}^{2}+(1-x)^{2}p_{j}^{2}-2(1-x^{2})p_{i}.p_{j}\eta_{i}\eta_{j}$ \\
	$p_{x}^{2}=(1+x)^{2}p_{i}^{2}+(1-x)^{2}p_{j}^{2}+2(1-x^{2})p_{i}.p_{j}$,\\
	next observe that for $\eta_{i} \eta_{j}=-1$ (ie one incoming and one outgoing particle),  $p_{x}^{'2}=p_{x}^{2}$, leading to cancellation of $\lambda^{2}$ term in $\mathcal{H}_{ij}$. In the other case when $\eta_{i}$ $\eta_{j}=1$ (both are either incoming or outgoing particles), changing $x\to 1/x$ leads to $p_{x}^{'2}\to p_{x}^{2}/x^{2}$ and the final result is again $\lambda^{2}$ independent.\footnote{Since $x\in(-1,1)$, changing it to $\frac{1}{x}$ leads to trouble at x=0. We have checked that the imaginary part of the quantity $B$ is nothing but the Coulomb/Sommerfeld factor. Since we have considered this term explicitly, we thus have taken only the real part of $B$ while evaluating the results.} This then explicitly verifies that the physical rate is independent of the IR regulator $\lambda$ that was introduced in the intermediate steps of the calculation, and is thus free of IR divergences. This also provides a crucial check on the calculation performed. \\
	To $\mathcal{O}(\alpha)$, the corrected double differential decay rate (with index $i$ being $0$ or $c$ for the neutral and charged $B$ decay mode) can be written as 
	\begin{equation}
	\frac{d^{2}\Gamma^{i}}{ds dq^{2}}=\frac{d^{2}\Gamma_{0}}{ds dq^{2}}\left(1 + \Delta^{i} \right) 
	\end{equation}
	where the correction factor $\Delta^{i}$ is defined as
	\begin{equation}
	\Delta^{i}=\frac{\left(\frac{d^{2}\Gamma^{i}}{ds dq^{2}} \right)}{\left(\frac{d^{2}\Gamma_{0}}{ds dq^{2}} \right)}-1
	\label{delta}
	\end{equation}
	$\Delta^{i}$ contains corrections due to infrared factors and non-infrared factors up to $\mathcal{O}(k)$ terms. We have explicitly checked that $\mathcal{O}(k^{2})$ piece is rather small, and therefore has been dropped from the analysis. The other relevant quantity is the shift in $R_{k^{\mu e}}$ due to the QED corrections. This shift, $\Delta_{R_{k^{\mu e}}}$, is defined as
	\begin{eqnarray}
	\Delta_{R_{k^{\mu e}}}^{i}=R_{k^{\mu e}}^{0}\left(\frac{\Delta \Gamma_{\mu}^{i}}{\Gamma_{\mu}^{i}}-\frac{\Delta \Gamma_{e}^{i}}{\Gamma_{e}^{i}}\right)  
	\label{deltaRK}
	\end{eqnarray}
	Below, we discuss the impact of these QED corrections. 
	
	\section{Results}
\begin{figure}[ht]
		\begin{subfigure}{.5\textwidth}
			\centering
			\includegraphics[scale=0.48]{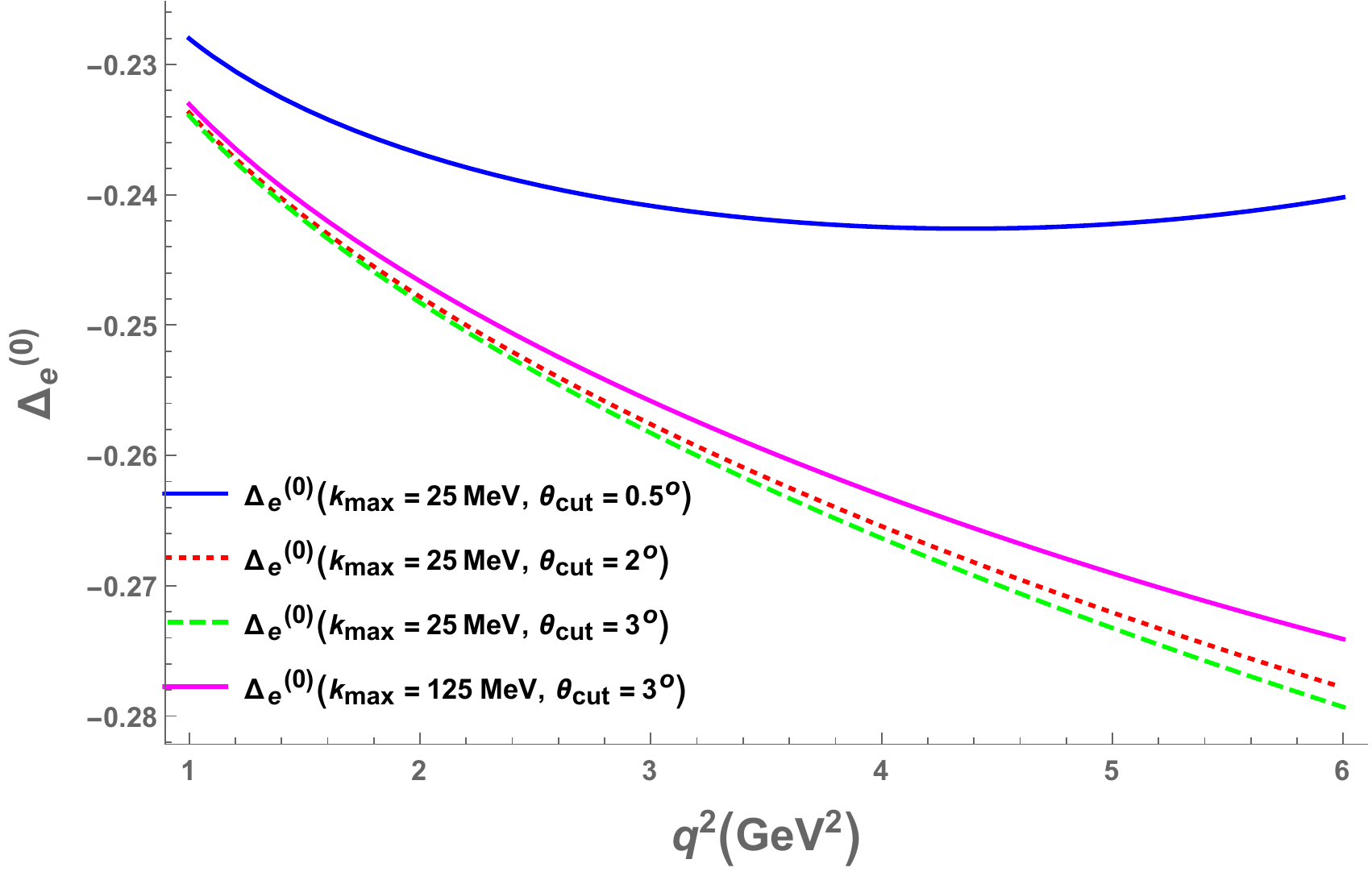}
			\caption{$\Delta^{(0)}_{e}$ vs. $q^{2}$}
		\end{subfigure}%
		\begin{subfigure}{.5\textwidth}
			\centering
			\includegraphics[scale=0.48]{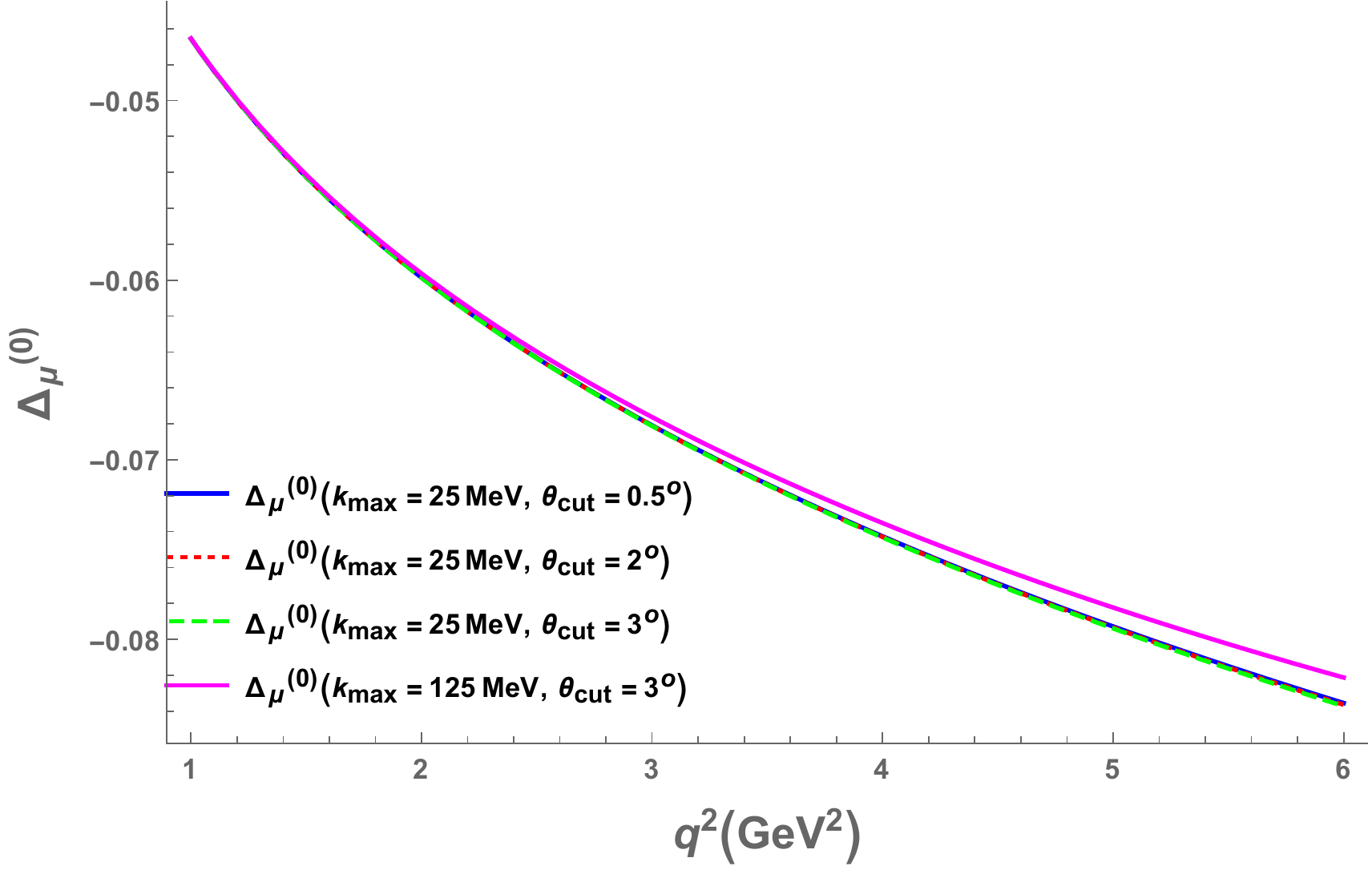}
			\caption{$\Delta^{(0)}_{\mu}$ vs. $q^{2}$}
		\end{subfigure}%
		\\\\
		\begin{subfigure}{.5\textwidth}
			\centering
			\includegraphics[scale=0.5]{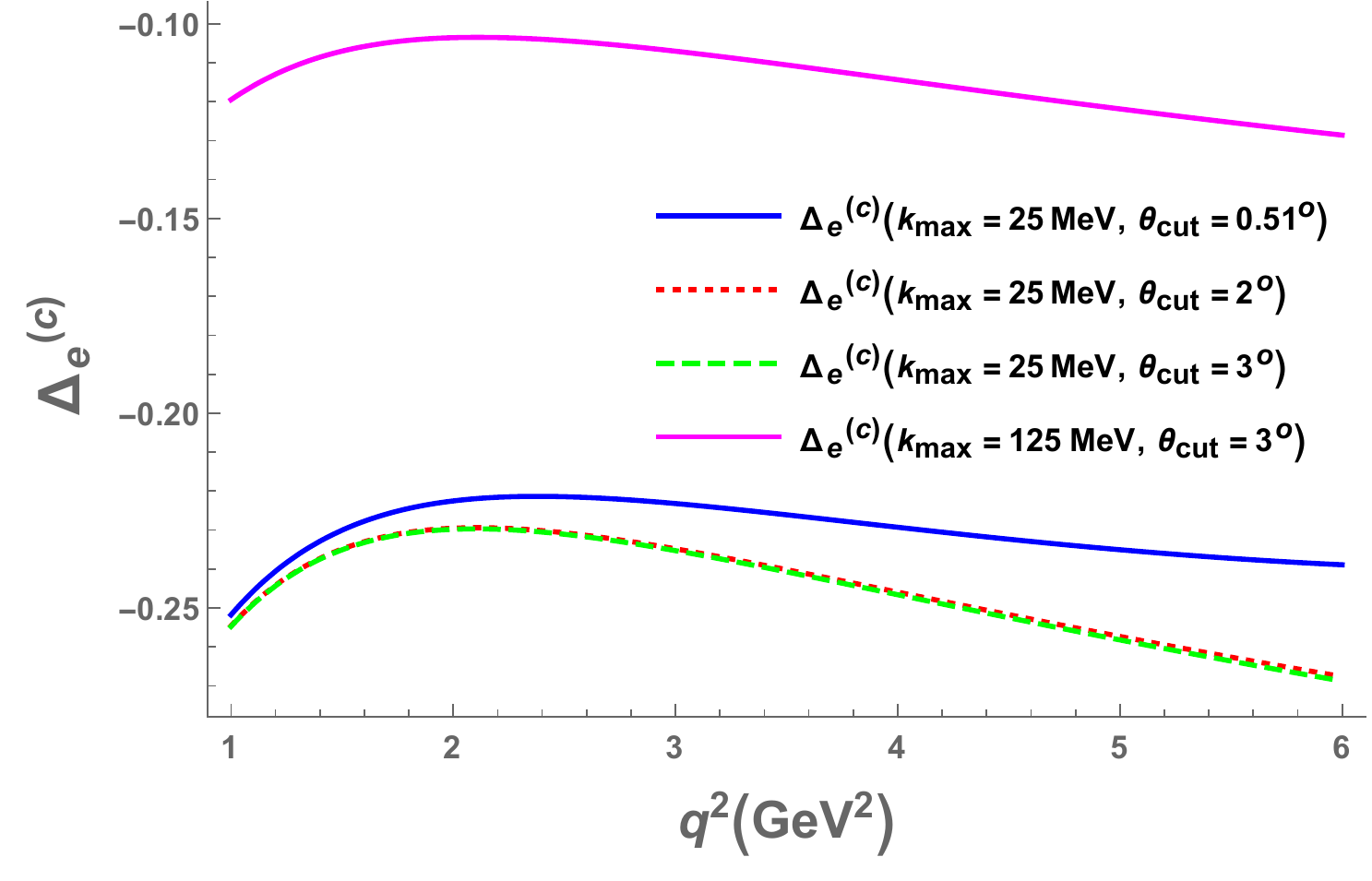}
			\caption{$\Delta_{e}^{(c)}$ vs. $q^{2}$}
		\end{subfigure}%
		\begin{subfigure}{.5\textwidth}
			\centering
			\includegraphics[scale=0.45]{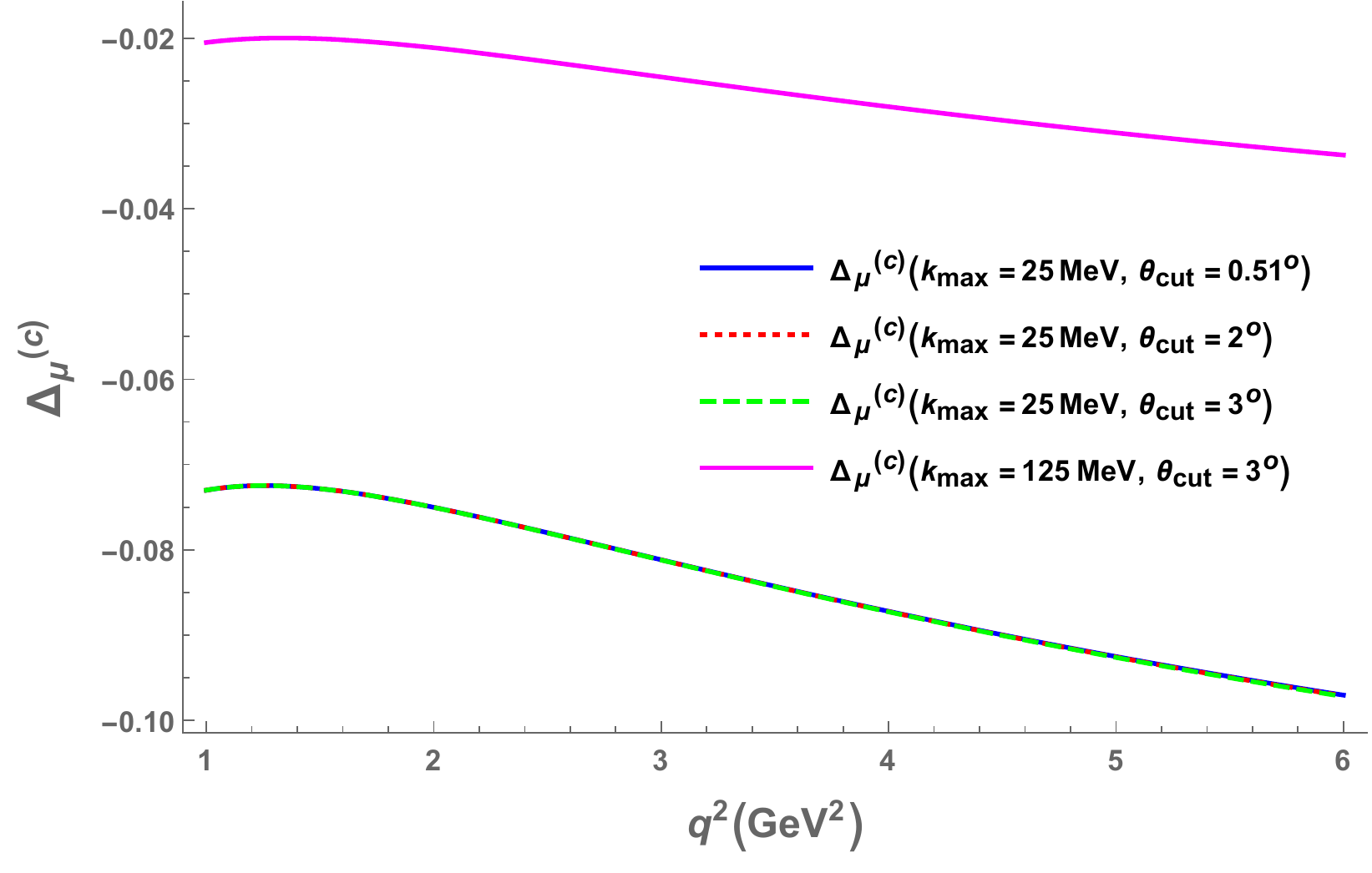}
			\caption{$\Delta_{\mu}^{(c)}$ vs. $q^{2}$}
		\end{subfigure}%
		\caption{$\mathcal{O}(\alpha)$ corrections to neutral and charged $B\to K\ell\ell$ modes. Left: electrons, Right: muons}
		\label{delta0C}
	\end{figure}

	To $\mathcal{O}(\alpha)$, the impact of QED corrections, real and virtual, is captured by the quantity $\Delta^i$ (see Eq.(\ref{delta})), with $i$ denoting the neutral ($i=0$) or charged ($i=c$) B decay. These are shown in Fig.(\ref{delta0C}) for the electron and muon channels. Shown in these figures are the correction factors for different choices of the maximum photon energy $k_{max}$ and the angular cut $\theta_{cut}$. First, it is to be noted that the correction factor for the electrons is about three times larger than that for the muons (and both are negative ie the correction factor will decrease the rate). This difference is essentially due to the smallness of electron mass compared to the muon mass by about two orders of magnitude. The QED corrections impact the more massive charged particles significantly less compared to lighter particles. 
	There is a mild dependence on the photon energy cut, $k_{max}$ chosen. The other important feature is the sensitivity to $\theta_{cut}$, particularly for the case of electrons. Choosing $\theta_{cut} \sim$ few degrees, this sensitivity essentially disappears. 
	
\begin{figure}[ht]
		\centering
		\includegraphics[scale=0.6]{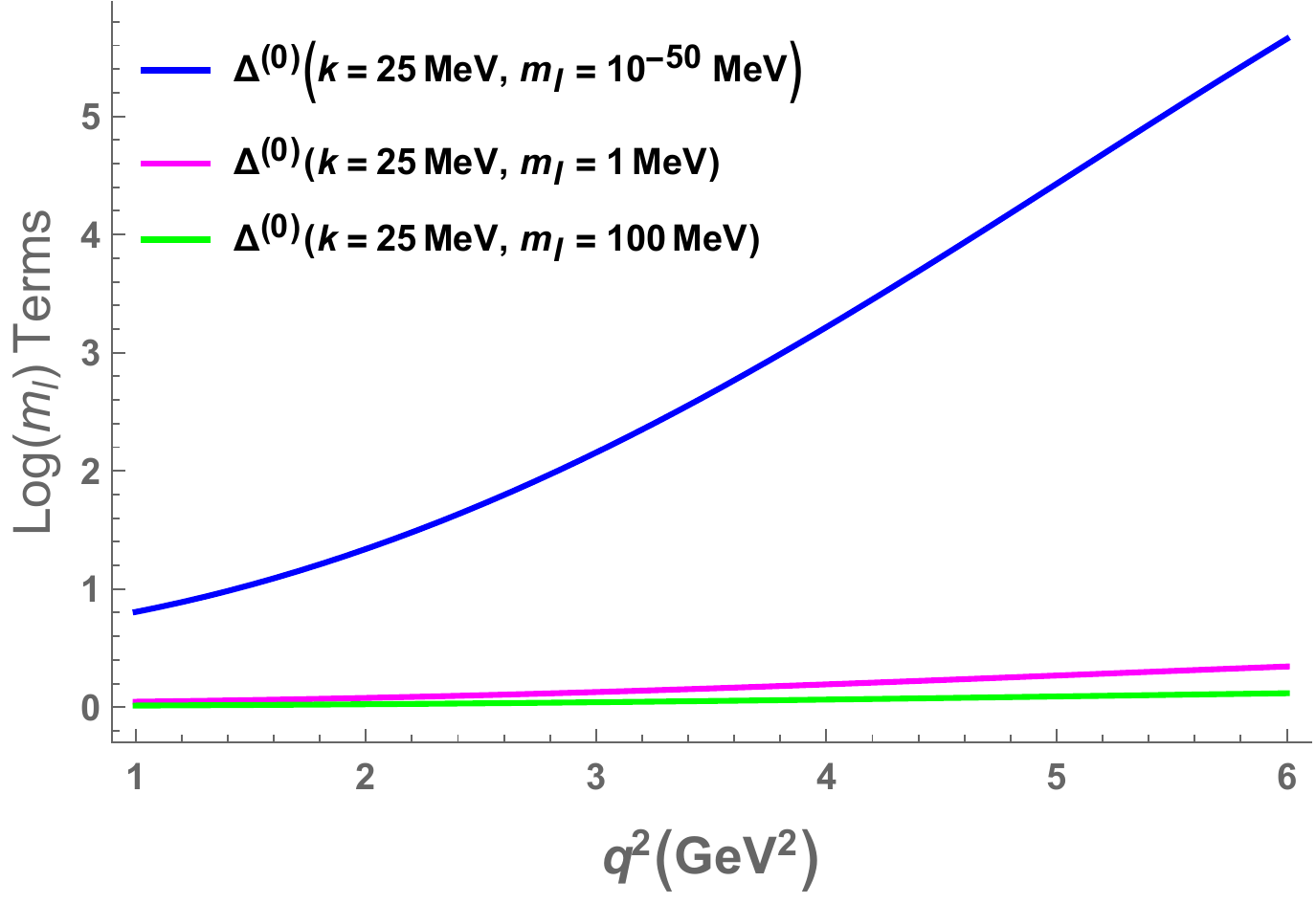}
		\caption{Behaviour of $\ln(m_{l})$ terms}
		\label{logml}
	\end{figure}

	An important set of terms are those in the IR terms that have logarithmic dependence on the lepton mass, $\ln(m_l)$. Fig.(\ref{logml}) shows the sensitivity on $m_l$. The lower two curves essentially are what is expected for the electron and muon case respectively (being about $10\%$ as in $\Delta^i$) while the one in blue is for the case when $m_l = 10^{-50}$ MeV. There is clearly a much larger contribution for this value and this is going to become larger as $m_l\to 0$. Employing $\theta_{cut} \sim$ few degrees takes care of this issue. These $\ln(m_l)$ terms correspond to hard collinear logs. These have been rigourously shown to cancel in \cite{Isidori:2020acz}. All other IR divergences, including the $\ln^2(m_l)$ terms are explicitly seen to cancel when the virtual corrections and real emission terms are added.\\
	
\begin{figure}[ht]
		\begin{subfigure}{.5\textwidth}
			\centering
			\includegraphics[scale=0.6]{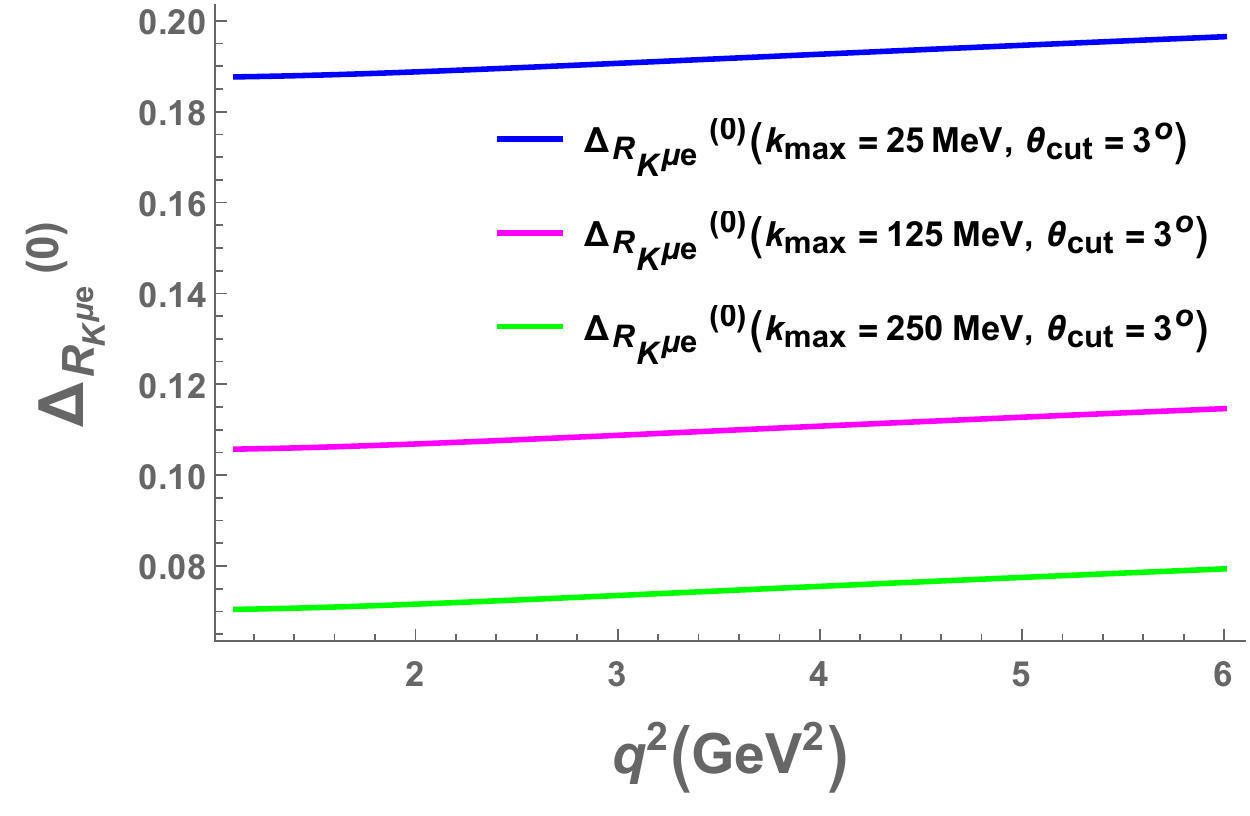}
			\caption{$\Delta^{0}_{R_{K}^{\mu e}}$ vs. $q^{2}$}
		\end{subfigure}%
		\begin{subfigure}{.5\textwidth}
			\centering
			\includegraphics[scale=0.5]{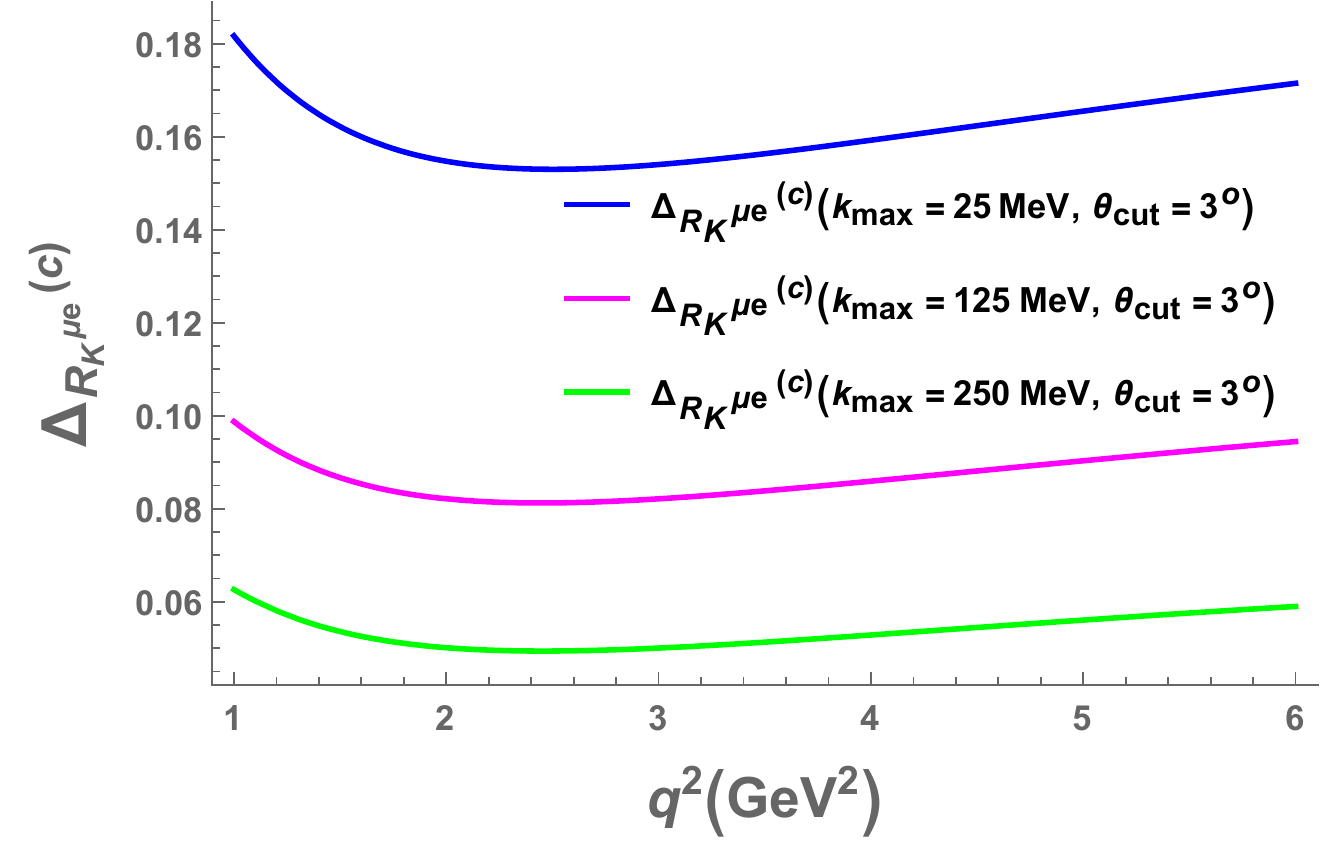}
			\caption{$\Delta_{R_{K}^{\mu e}}^{(c)}$ vs. $q^{2}$}
		\end{subfigure}%
		\caption{Shift in $R_{K}^{\mu e}$ due to $\mathcal{O}(\alpha)$ QED effects}
		\label{deltaRKplot}
	\end{figure}

	Fig.(\ref{deltaRKplot})	shows the impact of QED effects on $\Delta^{i}_{R_{K}^{\mu e}}$, defined in Eq.(\ref{deltaRK}) for $\theta_{cut} = 3^{\circ}$ as a function of $q^2$. Also, the charged mode is affected more as there are extra contributions from the contact term, which being proportional to $(Q_B+Q_K)$ are absent for the neutral mode. As the QED effects are sizable $\sim 20\%$ for the electrons ($\sim 5\%$ for muons, and both are negative), $\Delta^{i}_{R_{K}^{\mu e}}$, and thus, $R_{K}^{\mu e}$, increases. However, as mentioned above, all the quantities are sensitive to $k_{max}$ and $\theta_{cut}$.  The shift in $R_{K}^{\mu e}$, over the $q^2$ range, is about $20\%$ for $k_{max}=25$ MeV while with the increase in $k_{max}$ to $125$ MeV, it decreases to about $10\%$. This is quite expected since with the increase in $k_{max}$, the muons also start to effectively behave as electrons ie $m_e,\,\,m_{\mu} << k_{max}$ and thus both are affected similarly. We have checked that for such a case $\Delta^{i}_{R_{K}^{\mu e}}$ is very close to zero.
	
	In particular, choosing $\theta_{cut} \sim$ few degrees and $k_{max} \sim 250$ MeV, leads to $\sim 5\%$, (positive) shift in $R_{K}^{\mu e}$:
	\begin{equation}
	\Delta_{R_{K}^{\mu e}}^{(c)}=5.34\%,\hspace{1cm} \Delta_{R_{K}^{\mu e}}^{(0)}=7.43\%
	\end{equation}

	The results obtained here are in general agreement with the ones reported in \cite{Isidori:2020acz}. The electron modes receive large QED corrections, $\mathcal{O}(20\%)$ where as the muon modes receive smaller corrections. We have also checked that choosing different $k_{max}$ values for the muons and electrons changes the shift in $R_{K}^{\mu e}$ such that the final value of $R_{K}^{\mu e}$, including the QED effects, deviates from unity by a few percent. This is again in broad agreement with \cite{Isidori:2020acz}. The two studies essentially differ in arriving at the contact term(s), leading to some differences in the numerical values in the end, and in dealing with the $\ln(m_l)$ terms and phase space. Despite these differences, it is encouraging to see very similar conclusions being reached for the physical quantities.\\
	
	\section{Summary and Conclusions}
	We have calculated the $\mathcal{O}(\alpha)$ QED effects to $B\to K\ell^+\ell^-$ modes. These corrections are negative and thus reduce the rates. Starting with the non-radiative amplitude, and demanding the gauge invariance of the one photon emission amplitude (treating the mesons as point particles and employing scalar QED to calculate the amplitude), the contact term is fixed. Both the real and virtual QED effects are then calculated. While calculating the virtual corrections due to the contact term, there are UV divergences which should cancel against similar divergences from of higher dimensions, in particular two photon contact terms. For the present, we take a more phenomenological view point of these and simply discard the divergences and retain the finite terms in the calculations. A fictitious photon mass,$\lambda$, is chosen as the IR regulator. The physical differential decay rate is shown to be independent of the regulator $\lambda$, thereby showing the cancellation of the soft divergences. The physical rate as well as the ratio of the rates into muons and electrons are sensitive to the choice of the max photon energy, $k_{max}$ and the photon angle with respect to the charged lepton $\theta_{cut}$. We have also discussed the potentially large $\ln(m_l)$ terms whose effect is cancelled by choosing $\theta_{cut} \sim$ few degrees. Electron channels are found to receive $\sim 10$-$20\%$ corrections while for muons the corrections for the same  $k_{max}$ and $\theta_{cut}$ are around $5\%$. For $k_{max} \sim 250$ MeV, the corrections to the lepton flavour universality ration, $R_{K}^{\mu e}$ are abut $5\,\%$. This seems like a large correction, particularly given the fact that this observable has been experimentally known to show deviations from unity (the value of $R_{K}^{\mu e}$ expected within the standard model without these QED effects). This would worsen the tension between theory and experiments. However, a word of caution is in order. This $\mathcal{O}(5\%)$ positive shift in $R_{K}^{\mu e}$ is obtained by choosing the same $k_{max}$ and $\theta_{cut}$ for both the electrons and muons. Changing these to suit the actual experimental cuts would change these numbers. The results presented here are in general agreement with those obtained in \cite{Isidori:2020acz} which appeared recently, though there are some differences in the two calculations. 
	
	In conclusion, we have shown that the QED effects to $B\to K\ell^+\ell^-$ are an important source of corrections and should be systematically included. While the individual rates, particularly for the electrons, do receive reasonable corrections, with appropriate cuts (suiting the experiments), observables like $R_{K}^{\mu e}$ may only receive very nominal shifts $\sim$ few $\%$, and these corrections depend on the cuts imposed. This clearly shows the need for utmost care while comparing the experimental results with the theoretical calculations. The present study, and also \cite{Isidori:2020acz}, leave open the question of left over UV divergences and computation of two photon contact terms (and other higher dimensional operators appropriate to the issue), calling for extra effort in this direction. This is rather important so as to have unambiguous comparison with the experiments, particularly given that observables like $R_{K}^{\mu e}$ are heralded as very clean probes of the standard model, and therefore of new physics beyond it.  
	\\ \\
	{\bf Acknowledgements}\\
	We thank Anshika Bansal for collaboration in the initial stages and many discussions.

	\begin{appendices}
		\section{Real photon emission}
		\begin{eqnarray}
		\tilde{B}&=&\frac{-1}{8\pi^{2}}\eta_{i}\eta_{j}\int_{0}^{k_{max}}\frac{d^{3}k}{(k^{2}+\lambda^{2})^{1/2}}\left(\frac{p_{i\mu}}{k.p_{i}}-\frac{p_{j\mu}}{k.p_{j}} \right)^{2} \nonumber \\
		&=& \frac{-1}{8\pi^{2}}\eta_{i}\eta_{j}\int_{0}^{k_{max}}\frac{d^{3}k}{(k^{2}+\lambda^{2})^{1/2}}\left[\frac{m_{i}^{2}}{(k.p_{i})^{2}}+\frac{m_{j}^{2}}{(k.p_{j})^{2}}-\frac{2p_{i}.p_{j}}{(k.p_{i})(k.p_{j})}\right]  
		\end{eqnarray}
		
		First integral
		\begin{equation}
		\int_{0}^{k_{max}}\frac{d^{3}k}{(k^{2}+\lambda^{2})^{1/2}}\frac{1}{(k.p_{i})^{2}}=2\pi\frac{1}{m_{i}^{2}}\ln\left( \frac{k_{max}^{2}m_{i}^{2}}{E_{i}^{2}\lambda^{2}}\right) 
		\end{equation}
		
		Second integral
		\begin{equation}
		\int_{0}^{k_{max}}\frac{d^{3}k}{(k^{2}+\lambda^{2})^{1/2}}\frac{1}{(k.p_{j})^{2}}=2\pi\frac{1}{m_{j}^{2}}\ln\left( \frac{k_{max}^{2}m_{j}^{2}}{E_{i}^{2}\lambda^{2}}\right) 
		\end{equation}
		
		Third integral
		\begin{eqnarray}
		\int_{0}^{k_{max}}\frac{d^{3}k}{(\vec{k}^{2}+\lambda^{2})^{1/2}}\frac{1}{(k.p_{i})(k.p_{j})}&=& 2\int_{0}^{k_{max}}\frac{d^{3}k}{(k^{2}+\lambda^{2})^{1/2}}\int_{-1}^{1} \frac{dx}{((1+x)(k.p_{i})+(1-x)(k.p_{j}))^{2}} \nonumber \\
		&=&\frac{1}{2}\int_{0}^{k_{max}}\frac{d^{3}k}{(k^{2}+\lambda^{2})^{1/2}}\int_{-1}^{1}\frac{dx}{k^{2}p_{x}^{2}}  \nonumber \\
		&=&\frac{1}{2}\int_{0}^{k_{max}}\frac{d^{3}k}{(k^{2}+\lambda^{2})^{1/2}}\int_{-1}^{1}\frac{dx}{\omega^{2}E_{x}^{2}-k^{2}p_{x}^{2}} \nonumber \\
		&=& 2\pi\frac{1}{2}\int_{-1}^{1}\frac{dx}{p_{x}^{2}}\ln\left( \frac{k_{max}^{2}p_{x}^{2}}{E_{x}^{2}\lambda^{2}}\right) 
		\end{eqnarray}
		Substituting back all the individual integrals,
		\begin{equation}
		\tilde{B_{ij}}=\frac{Q_{i}Q_{j}\eta_{i}\eta_{j}}{2\pi}\left\lbrace \ln\left( \frac{k_{max}^{2}m_{i}m_{j}}{\lambda^{2}E_{i}E_{j}}\right)-\frac{p_{i}.p_{j}}{2}\left[ \int_{-1}^{1}\frac{dx}{p_{x}^{2}}\ln\left(\frac{k_{max}^{2}}{E_{x}^{2}} \right)+\int_{-1}^{1}\frac{dx}{p_{x}^{2}}\ln\left(\frac{p_{x}^{2}}{\lambda^{2}} \right) \right]  \right\rbrace
		\end{equation}
		
		\section{Virtual photon corrections}
		\begin{align*}
		B&=\frac{-i}{8\pi^{3}}\eta_{i}\eta_{j}\int \frac{d^{4}k}{k^{2}-\lambda^{2}}\left(\frac{2p_{i}\eta_{i}-k}{k^{2}-2k.p_{i}\eta_{i}}+ \frac{2p_{j}\eta_{j}+k}{k^{2}+2k.p_{j}\eta_{j}}\right)^{2}\\
		&=\frac{-i}{8\pi^{3}}\eta_{i}\eta_{j}\int \frac{d^{4}k}{k^{2}-\lambda^{2}}\left[\left( \frac{2p_{i}\eta_{i}}{k^{2}-2k.p_{i}\eta_{i}}+\frac{2p_{j}\eta_{j}}{k^{2}-2k.p_{j}\eta_{j}}\right)^{2}-k^{2}\left( \frac{1}{k^{2}-2k.p_{i}\eta_{i}}-\frac{1}{k^{2}-2k.p_{j}\eta_{j}}\right)^{2}  \right]  \\
		&=\frac{-i}{8\pi^{3}}\eta_{i}\eta_{j}\left[ \int \frac{d^{4}k}{k^{2}-\lambda^{2}}\left( \frac{4m_{i}^{2}}{(k^{2}-2k.p_{i}\eta_{i})^{2}}+\frac{4m_{j}^{2}}{(k^{2}+2k.p_{j}\eta_{j})^{2}}+\frac{8p_{i}.p_{j}\eta_{i}\eta_{j}}{(k^{2}-2k.p_{i}\eta_{i})(k^{2}+2k.p_{j}\eta_{j})}\right)\right. \\&\left. -\int d^{4}k\left(\frac{1}{(k^{2}-2k.p_{i}\eta_{i})^{2}}+\frac{1}{(k^{2}+2k.p_{j}\eta_{j})^{2}}-\frac{2}{(k^{2}-2k.p_{i}\eta_{i})(k^{2}-2k.p_{j}\eta_{j})} \right)  \right] 
		\end{align*}

		First integral
		\begin{equation}
		\int \frac{d^{4}k}{(k^{2}-\lambda^{2})}\frac{1}{(k^{2}-2k.p_{j}\eta_{j})^{2}}=\frac{-i\pi^{2}}{2m_{j}^{2}}\ln\left(\frac{m_{j}^{2}}{\lambda^{2}} \right) 
		\end{equation}
		
		Second integral
		\begin{equation}
		\int \frac{d^{4}k}{(k^{2}-\lambda^{2})}\frac{1}{(k^{2}-2k.p_{i}\eta_{i})^{2}}=\frac{-i\pi^{2}}{2m_{i}^{2}}\ln\left(\frac{m_{i}^{2}}{\lambda^{2}} \right) 
		\end{equation}
		
		Third integral 
		\begin{eqnarray}
		\int \frac{d^{4}k}{(k^{2}-\lambda^{2})}\frac{1}{(k^{2}-2k.p_{i}\eta_{i})(k^{2}-2k.p_{j}\eta_{j})}&=&\frac{-i\pi^{2}}{4}\int_{-1}^{1}\frac{dx}{p_{x}^{'2}}\ln\left(\frac{\lambda^{2}+p_{x}^{'2}}{\lambda^{2}} \right) \nonumber \\  &\xrightarrow{\lambda\to 0}& \frac{-i\pi^{2}}{4}\int_{-1}^{1}\frac{dx}{p_{x}^{'2}}\ln\left(\frac{p_{x}^{'2}}{\lambda^{2}} \right) 
		\end{eqnarray}
		
		Fourth and fifth integral
		\begin{equation}
		\int d^{4}k\frac{1}{(k^{2}-2k.p_{j}\eta_{j})^{2}}=-i\pi^{2}\ln\left(m_{j}^{2} \right) 
		\end{equation}
		
		Sixth integral 
		\begin{equation}
		\int d^{4}k\frac{1}{(k^{2}-2k.p_{i}\eta_{i})(k^{2}-2k.p_{j}\eta_{j})}=\frac{-i\pi^{2}}{2}\int_{-1}^{1}dx\ln\left(p_{x}^{'2} \right) 
		\end{equation}
		
		Substituting back all the individual integrals,  
		\begin{equation}
		B=\frac{-\eta_{i}\eta_{j}}{2\pi}\left[ \ln\left( \frac{m_{i}m_{j}}{\lambda^{2}}\right)+\frac{1}{4}\int_{-1}^{1}dx \ln\left( \frac{p_{x}^{'2}}{m_{i}m_{j}}\right)+\frac{p_{i}.p_{j}\eta_{i}\eta_{j}}{2}\int_{-1}^{1}\frac{dx}{p_{x}^{'2}}\ln\left( \frac{p_{x}^{'2}}{\lambda^{2}}\right)   \right] 
		\end{equation}
	\end{appendices}
	\bibliographystyle{unsrt}
	\bibliography{radiative_correction}

	\end{document}